\documentclass{ieeetj}
\usepackage{cite}
\usepackage{amsmath,amssymb,amsfonts}
\usepackage{algorithmic}
\usepackage{graphicx,color}
\usepackage{rotate}
\usepackage{textcomp}
\usepackage{xcolor}
\usepackage{hyperref}
\hypersetup{hidelinks=true}
\usepackage{algorithm,algorithmic}
\def\BibTeX{{\rm B\kern-.05em{\sc i\kern-.025em b}\kern-.08em
    T\kern-.1667em\lower.7ex\hbox{E}\kern-.125emX}}
\AtBeginDocument{\definecolor{tmlcncolor}{cmyk}{0.93,0.59,0.15,0.02}\definecolor{NavyBlue}{RGB}{0,86,125}}

\makeatletter
\renewcommand{\p@subsection}{\thesection-}
\makeatother

\def\OJlogo{\vspace{-4pt}$<$Society logo(s) and publication title will appear here.$>$}
\def\seclogo{\vspace{10pt}$<$Society logo(s) and publication title will appear here.$>$}

\def\authorrefmark#1{\ensuremath{^{\textbf{#1}}}}

\usepackage{booktabs}
\usepackage{multirow}
\usepackage{caption}
\usepackage{subcaption}
\usepackage{siunitx}
\usepackage{float}

\usepackage[switch]{lineno}

\begin{document}

\receiveddate{13 May, 2026}
\accepteddate{22 June, 2026}
\doiinfo{/OJCAS.2026.3714974}
\currentdate{16 July 2026}

\title{The SpiNNaker2 chip: a many-core platform for flexible and scalable brain-inspired computing}

\author{
Stefan Scholze\authorrefmark{1,3$^\dagger$},
Johannes Partzsch\authorrefmark{1,2$^\dagger$},
Sebastian Höppner\authorrefmark{1$^\dagger$},
Florian Kelber\authorrefmark{1$^\dagger$},
Andreas Dixius\authorrefmark{1$^\dagger$},
Marco Stolba\authorrefmark{1$^\dagger$},
Sirine Arfa\authorrefmark{1},
Marc Berthel\authorrefmark{1},
Georg Ellguth\authorrefmark{1},
Jim Garside\authorrefmark{4},
Hector A. Gonzalez\authorrefmark{1,3},
Stephan Hartmann\authorrefmark{1},
Thomas Kiel-Hocker\authorrefmark{1},
Dongwei Hu\authorrefmark{4},
Matthias Jobst\authorrefmark{1,2}, 
Khaleelulla Khan Nazeer\authorrefmark{1,3},
Tim Langer\authorrefmark{1,2},
Chen Liu\authorrefmark{1},
Gengting Liu\authorrefmark{4},
Matthias Lohrmann\authorrefmark{1},
Mantas Mikaitis\authorrefmark{4,5}, 
Felix Neumärker\authorrefmark{1},
Amirhossein Rostami\authorrefmark{1,3},
Stefan Schiefer\authorrefmark{1,3},
Tilo Schubert\authorrefmark{1},
Delong Shang\authorrefmark{4},
Bernhard Vogginger\authorrefmark{1,3},
Yexin Yan\authorrefmark{1,3},
Steve Furber\authorrefmark{1},
and Christian Mayr\authorrefmark{1,2,3}
}

\affil{Chair of Highly-Parallel VLSI-Systems and Neuro-Microelectronics, Technische Universität Dresden, Germany}
\affil{Centre for Tactile Internet with Human-in-the-loop (CeTI)}
\affil{Center for Scalable Data Analytics and Artificial Intelligence (ScaDS.AI Dresden)}
\affil{Advanced Processor Technologies Group, Department of Computer Science, University of Manchester, UK}
\affil{now with School of Computer Science, University of Leeds, Leeds, UK\\
$^\dagger$These authors contributed equally to this work.
}

\corresp{Corresponding author: Johannes Partzsch (email: johannes.partzsch@tu-dresden.de).}

\authornote{
This work is partly funded by the German Federal Ministry of Research, Technology and Space (BMFTR) in DAAD project 57616814 (SECAI), by BMFTR and the Free State of Saxony within ScaDS.AI center of excellence for AI research, by the German Federal Ministry for Economic Affairs and Energy (BMWE) under contract 01MN23004F (ESCADE), and by DFG (German Research Foundation) as part of Germany’s Excellence Strategy – EXC 2050/1 and EXC 2050/2 – Project ID 390696704 – Cluster of Excellence “CeTI” of TU Dresden.
This work received funding from EFRE (European Regional Development Fund) and the Free State of Saxony under grant 100373652 (SpiNNcloud).
The work was funded by the EC Horizon 2020 Framework Programme under grant agreements 720270 (HBP SGA1), 785907 (HBP SGA2), 945539 (HBP SGA3) and Horizon Europe grant agreements 101147319 (EBRAINS 2.0) and 101120727 (PRIMI). 
The authors gratefully acknowledge the computing time on the high-performance computer at the NHR Center of TU Dresden. This center is jointly supported by BMFTR and the state governments participating in the NHR (www.nhr-verein.de/unsere-partner).
}

\begin{abstract}
In deep learning, efficiency gets more and more important to compensate for the ongoing growth in model sizes and applications.
Neuromorphic hardware has long been advocated as an upcoming alternative to deep networks, taking inspiration from the brain for achieving unprecedented energy efficiency.
However, demonstrations of these gains only recently began to grow in complexity and real-world applicability.
With SpiNNaker2, we present a chip that bridges the gap between deep networks and neuromorphic computing and allows for flexible exploration of computing approaches that combine both worlds.
It features 152 processing elements equipped with an ARM M4F processor and dedicated accelerators, an extended SpiNNaker routing fabric for scalable event-based communication and a range of external interfaces for system integration, including Gbit Ethernet and an LPDDR4 memory interface.
We demonstrate performance and efficiency of the SpiNNaker2 chip for neuromorphic and deep network workloads, as well as novel event-based computing approaches. 
For deep network workloads, the chip achieves up to 4.5 TOPS in high performance mode and up to 2.7 TOPS/W efficiency in high efficiency mode for INT8 workloads. 
The chip supports spiking neural networks with \num{>150000} neurons and $>1.8$ billion synaptic events/s when simulated with a 1 ms time step.
Its low baseline power of less than 250 mW allows for efficiency even under varying workload conditions, allowing to explore sparse and event-based modes of computation.
All this demonstrates the chip's capabilities as a universal hardware platform for scalable brain-inspired computing and its combinations with mainstream deep network approaches.\vspace{-0.2cm}\\
\end{abstract}

\maketitle

\section{Introduction}

The power draw caused by artificial intelligence (AI) applications is growing drastically due to increased usage demands.
To counterbalance this trend, more power-efficient hardware platforms are needed.
Neuromorphic computing has been promising a boost in energy efficiency by taking over working principles of the brain.
Those include event-based computation, locality, and co-location of memory and processing.
However, demonstrations of neuromorphic hardware are still lacking behind mainstream AI \cite{ostrau2022benchmarking,vogginger2024datacenter}.

On the other hand, recent developments in machine learning for more efficient processing, such as million mixture-of-experts in large language models \cite{belcak2023language,he2024mixture} or state-space models \cite{schoene2024scalable}, move towards similar modes of operation as used by neuromorphic computing.
GPUs are not well suited for executing such types of algorithms due to the required fine-grained parallelism and irregular workload patterns. Even novel deep learning architectures are not designed to leverage such dynamically sparse computation \cite{SambaNovaArch,TenstorrentArch,reuther_lincoln_2023}.

In parallel, the field of neuromorphic computing is reaching a plateau of maturity. Dedicated neuromorphic hardware systems, implementing spiking neural networks (SNN) have been significantly advanced in terms of size \cite{schmitt2017hwloop}, efficiency \cite{frenkel2022reckon,frenkel2023review,shrestha2024loihi2,yao2024speck} and usability \cite{davies2021loihi}.
Moreover, platforms for hybrid execution of spiking and artificial neural networks have been developed and successfully deployed \cite{tang2023seneca,pei2019tianjic,ma2022tianjicx}.

Among the larger-scale neuromorphic systems, SpiNNaker \cite{furber2014spinnaker} stands out with its focus on scalable event communication \cite{furber2013spinnarch} and its general programmability. Unlike other neuromorphic hardware whose architecture follows a synapse-neuron processing scheme, SpiNNaker is completely agnostic to the algorithm or neural network model by realizing them in software on microcontroller cores.

In this paper, we present the SpiNNaker2 chip, which follows the concept of the established SpiNNaker system, but significantly extends it in its computational capabilities and circuit support for efficient realization of event-based algorithms.
Compared to the SpiNNaker chip, the number of processing elements (cores) is significantly increased from 18 to 152. Each core now 
includes dedicated accelerators for often-used functions, such as matrix multiplications, convolutions, as well as exponentials and logarithms. The chip can adapt flexibly to local workload fluctuations in spiking and event-based neural networks by dynamic voltage and frequency scaling (DVFS) of processing elements \cite{hoeppner2019dynamic}. This adaptability is further supported by a low-leakage design implementation strategy, resulting in a low baseline power. With an extended version of the SpiNNaker routing fabric \cite{furber2013spinnarch} and power-adaptive chip-to-chip links optimized for event-based communication, the SpiNNaker2 chip can be used to build scalable neuromorphic compute platforms up to 10 million cores \cite{mayr2019spinnaker}.

In the following, we present the design and implementation approach for the SpiNNaker2 chip and introduce its architecture, giving details on the individual components and the design implementation. We present a range of spiking and deep neural networks, as well as new event-based algorithms, demonstrating the versatility of the chip.

\section{Design and Implementation Approach}

\subsection{Design Philosophy}

The SpiNNaker2 chip follows key architectural assumptions of the SpiNNaker system \cite{furber2014spinnaker}, further developing and enhancing them for widening the application focus and improving system efficiency.
Its core principles are:
\begin{itemize}
    \item {\bf Flexibility} by implementation of algorithms predominantly in software on microcontrollers. This makes SpiNNaker2 much more versatile than systems dedicated to a class of algorithms, such as those for spiking or deep neural networks. 
    \item {\bf Distributed processing} on comparatively tiny, independently operating processing elements (PE). This allows for integration of many PEs on a chip and flexible distribution of heterogeneous workloads.
    \item {\bf Event-based communication} together with matching event-based algorithms offers the potential for significantly reduced data transfer. Motivated by spiking neural networks in SpiNNaker, the concept is extended to events with flexible payloads for supporting a wide range of event-based algorithms.
    \item {\bf Workload adaptation} for highest energy efficiency, by offering a low baseline power and different performance modes through dynamic voltage and frequency scaling (DVFS). With that, algorithmic savings, e.g. in compute, memory access and communication, can be directly translated into energy savings.
    \item {\bf Hardware acceleration} of the most common operations in neural network based algorithms.
    \item {\bf Robust design} by employing established digital design implementation techniques, focusing on reliable operation of all system components under the full range of process, voltage and temperature variations. This is a key prerequisite for a usable large-scale system.
\end{itemize}

The core focus of the SpiNNaker2 chip design is to provide a platform for hardware realization of different flavors and combinations of neural networks, with the potential to accommodate other classes of algorithms that employ sparse, event-based communication and computation.

\subsection{Scalability of the System}

\begin{figure}
    \centering
    \includegraphics[width=0.9\linewidth]{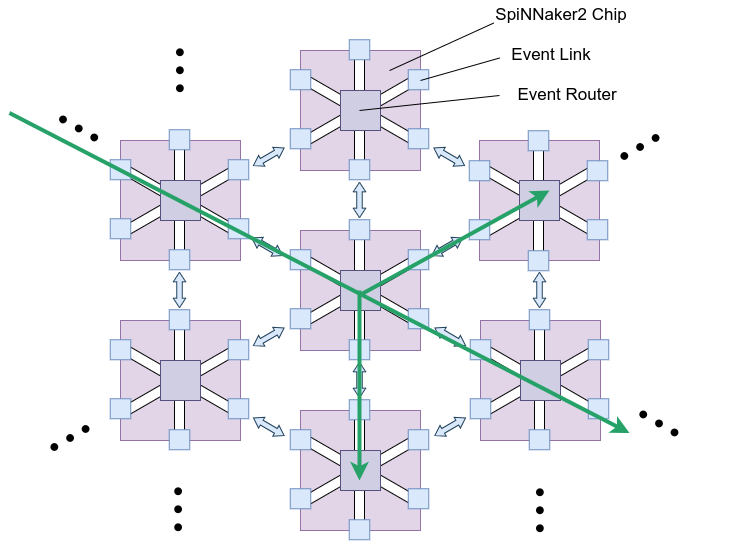}
    \caption{Hexagonal SpiNNaker communication fabric. The green lines denote an example route of a source address.}
    \label{fig:s2hexgrid}
\end{figure}

The SpiNNaker2 chip is designed to be part of a scalable system that can consist of a few chips to many thousands, allowing like SpiNNaker to implement both mobile and mainframe systems. 
The SpiNNaker routing fabric \cite{furber2013spinnarch} is the core of this system architecture.
It is fitted to neural connectivity in the brain, where one source neuron is connected to many destinations.
Each chip is representing a node in a hexagonal grid.
The on-chip event router allows for multicasting packets in each node, allowing for an efficient tree-like realization of brain connectivity.
Fig.~\ref{fig:s2hexgrid} illustrates the routing fabric and an example multi-cast connection.
The hexagonal grid could be arbitrarily extended, limited only by the maximum acceptable latency of the system.
Messages in the SpiNNaker2 routing fabric carry small payloads or no payload at all, in line with the communication needs of event-based and spiking neural networks.
Those types of algorithms profit from the flexibility and reduced communication load of small packets and a minimal protocol overhead.

\section{Chip Architecture}

\begin{figure}
    \centering
    \includegraphics[width=0.9\columnwidth]{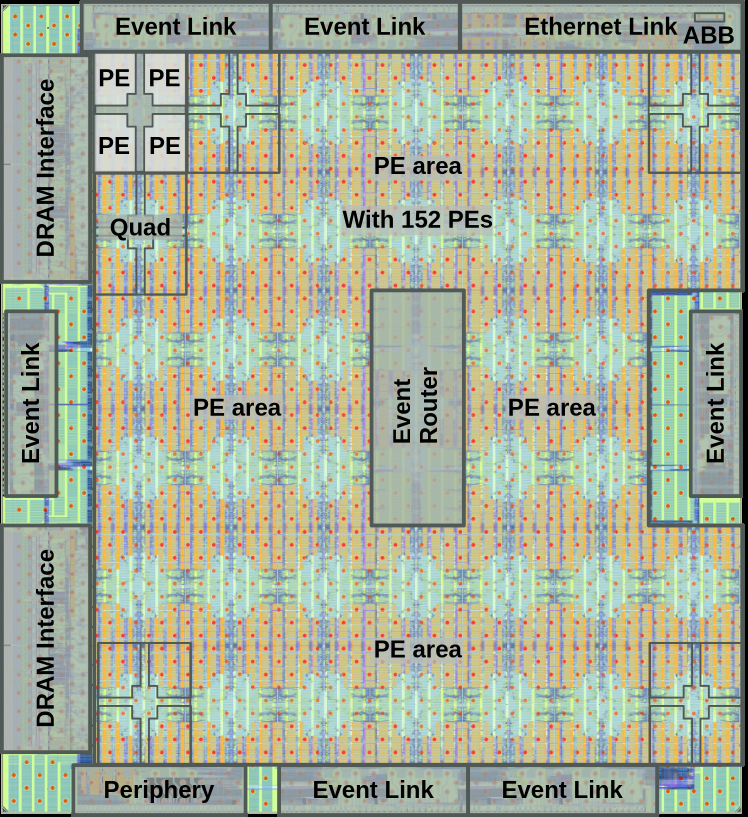}
    \caption{SpiNNaker2 chip layout with components}
    \label{fig:s2layout}
\end{figure}

\subsection{Overview}
The layout view of the SpiNNaker2 chip is shown in Fig.~\ref{fig:s2layout}. The main computation component is the quad unit (Quad), each containing four processing elements (PEs). Event-based communication is realized by the event router, located in the center of the chip for minimized latency from all PEs. There are communication interfaces to neighboring chips for different ranges. Long-range and short-range communication components are located on each edge of the chip for the hexagonal communication grid. Two DRAM interfaces are located on the left side of the chip. An Ethernet link is provided for high-speed configuration, located on the top right. General purpose interfaces for access to sensors, robots or debugging devices are located in the top left corner in a periphery module.
Within the periphery module a processor (ARM M4F) is located with full access to the chip's memory and configuration bits, used for various control and background tasks. The periphery unit further contains a single ADPLL clock multiplier used for central frequency multiplication, true random number generation and the adaptive body bias generator as further described in Sec. \ref{subsubsec:power_management}.
For addressing reasons but also to ease chip size scalability, we used a tile based implementation and placement flow, which can be seen in the uniform area and macro organization of the chip.
The components, the power-management and the design implementation details are explained in the following sections.

\subsection{Components}

\subsubsection{Processing Element (PE)}
\label{sec:pe}

\begin{figure}
    \centering
    \includegraphics[width=0.7\linewidth]{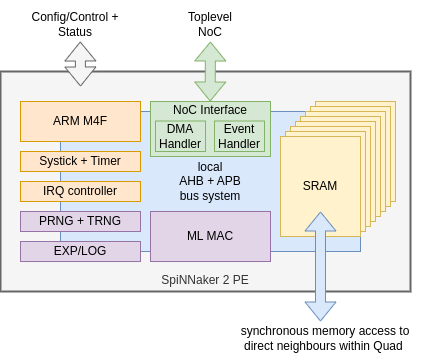}
    \caption{SpiNNaker2 processing element (PE) components}
    \label{fig:s2pe}
\end{figure}

The PE is the basic processing unit (see Fig.~\ref{fig:s2pe}). It features an ARM M4F core for general purpose execution and accelerators for specific computations. The PE contains 128 kByte SRAM memory for program code and data, which can also be accessed over the Network-on-Chip (NoC, see section \ref{subsubsec:noc}). Accelerators and NoC interface can run independently of each other. All parts of the PE are interconnected via an AHB/APB bus system and further controlled through 45 different interrupts evaluated by the ARM core. Accelerators are memory-mapped into a joint address space for easy programming from the ARM core. Each PE has a fine-grained DVFS functionality implemented (see \ref{subsubsec:power_management}). Three different timers are available for applications and time measurements, supplied with a 1 MHz reference clock.
 
\underline{Numerical Accelerator:} Each PE implements an iterative exponential/natural logarithm accelerator\cite{mikaitis_explog_2018}. Applied are s16.15 and s0.31 fixed-point and single-precision floating-point formats for both operand and result independently. For the floating-point format, subnormals and out-of-range notifiers (+/-infinity) are supported. For fixed-point formats, the output is saturated to a maximum and minimum value. The precision is configurable by user-programming between 1 and 16 iterations, taking 7 to 22 clock cycles per operation. The module achieves an accuracy between 1-2 units-in-the-last-place (ulp), as well as monotonicity. 

\underline{Random Number Generators:} For processes dependent on entropy, the SpiNNaker2 architecture includes Pseudo Random Number Generators (PRNG) in each PE, supplying 32bit per clock cycle, and one global True Random Number Generator (TRNG) \cite{neumaerker_rng_2016}. The PRNGs implement the MARS KISS64 algorithm. The TRNG uses a phase frequency detection signal from multiple ADPLL clock generators as a randomization source. The TRNG output can directly be accessed, or sent to the PRNG modules for post-processing. Entropy sources can also be used to scramble the PRNGs.

\underline{Rounding Accelerator:} Each PE is equipped with acceleration of signed/unsigned integer rounding and saturation of 64-, 32- and 16-bit values \cite{mikaitis2021}. Floating-point input can be rounded to BFloat16 output. The available modes are round-to-nearest with up rounding on a tie and stochastic rounding by applying random numbers from the PRNG. The rounding bit position can be configured to round any number of least-significant bits up to 32. Four threads work in parallel supplied by separate PRNG channels, each generating output in 3–4 cycles.

\underline{Machine Learning Accelerator (MLA):} To support fast and energy efficient execution of DNN inference, each PE provides an accelerator for 8bit/16bit signed/unsigned execution of 2D matrix multiplication and 2D convolution \cite{alipaper, kelberMappingDeepNeural2020}. The accelerator consists of a 16x4 output-stationary multiply-and-accumulate (MAC) array and 4 post-processing modules. Each 2 neighboring 8bit MAC cells can be fused together to accept 16bit input column- or row-wise and accumulated in a 24bit register. The output can be quantized via shift and truncation in a post-processing block and written out as 8bit, 16bit or 32bit. Data can be fetched directly over the NoC with prefetch. Further data reuse is achieved by a shift register for the input during 2D convolution. 

\underline{NoC Interface:} All PEs can request or send out NoC packet streams in parallel with a DMA module. NoC packets can be simple read/write requests/responses, control packets, exception packets, protocol packets or SpiNNaker2 router packets (see section \ref{subsubsec:event_router}). 
The event handler processes received SpiNNaker2 packets with two configurable filters and a default handler. It can filter multicast spike packets and store the 32-bit keys into a FIFO buffer mapped to the SRAM.
This allows to process received spikes in a batch, e.g. when a certain fill-level of the FIFO is reached. Non-filtered packets are processed by the default packet handler.
See supplement for further details about event handler.

\subsubsection{Quad unit}

The chip's PEs are physically and logically grouped into a Quad.
Each Quad is responsible for clock and power management of its
4 PEs and provides local management features (hardware semaphores, interrupt requests).
PEs can be operated at one of 4 performance levels, a predefined combination of one of two supply voltages and a clock frequency.
Frequency scaling is realized by local frequency division of the centrally-generated ADPLL clock.
PEs operate in globally asynchronous, locally synchronous (GALS) mode,
either individually or as a locally coupled synchronous group within a Quad.
PEs within the same Quad can access the full Quad SRAM memory if in synchronous mode.

\subsubsection{Network on Chip (NoC)} \label{subsubsec:noc}
\begin{figure}
    \centering
    \includegraphics[width=0.75\linewidth]{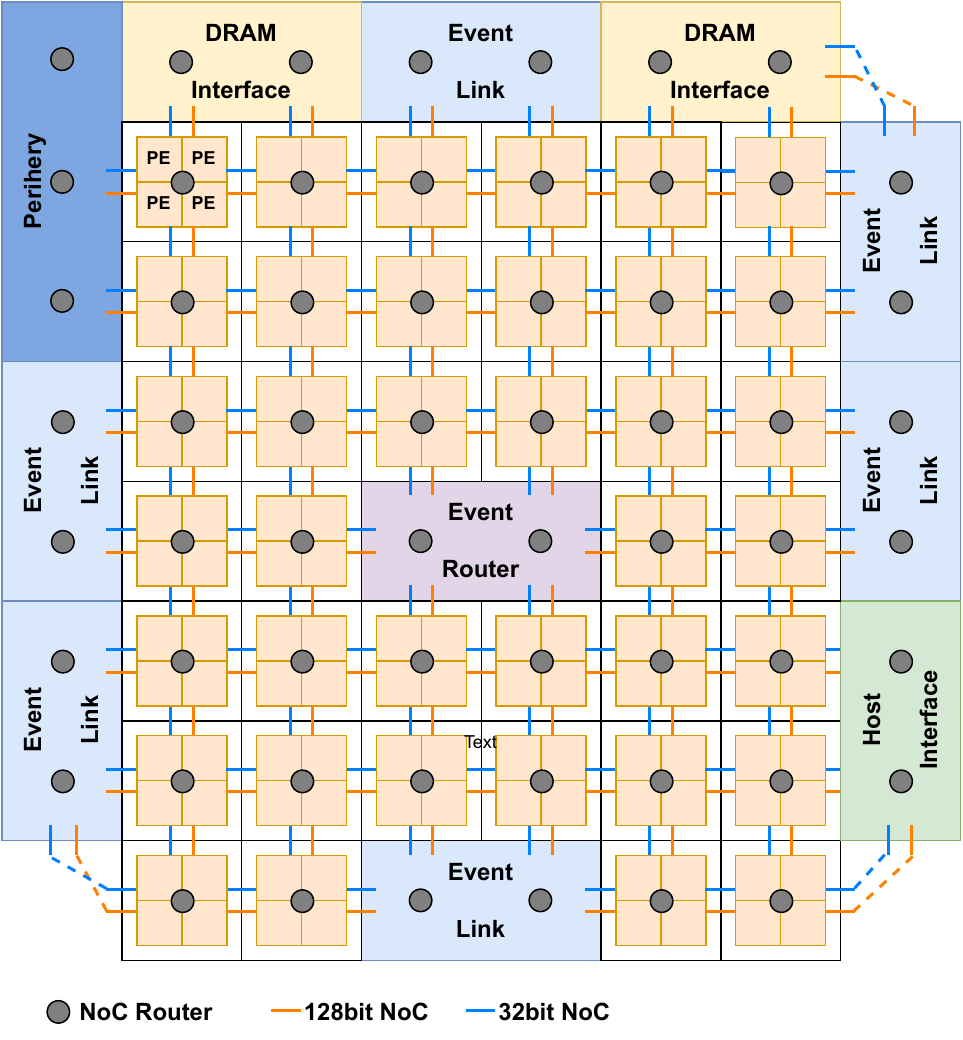}
    \caption{SpiNNaker2 NoC connections}
    \label{fig:s2noc}
\end{figure}

The NoC serves as the central communication infrastructure, 
connecting all on-chip components in a 2D mesh topology. It comprises two distinct, interconnected networks: 
the Data NoC (DNoC) and the Configuration NoC (CNoC) see Fig.~\ref{fig:s2noc}.
The DNoC handles large data transfers, such as DMA operations and spike packets. 
With a flit size of 192 bits, it enables the transfer of 128 bit data packets in a single transaction, ensuring high throughput. 
Operating at \SI{300}{\mega\hertz}, the DNoC achieves a bisection bandwidth of \SI{307.2}{\giga\bit/\second}.
Following the GALS design, the clocks of the PEs and NoC routers are asynchronous, interfaced via asynchronous FIFOs. This synchronization introduces a latency of 5 clock cycles per hop.
The CNoC is primarily designed for configuration and production testing. 
It operates with a 32-bit flit size and shares the same packet format as the DNoC. Running at a 100 MHz reference clock, it functions independently of the ADPLL clock.
Each Quad unit includes both a DNoC and CNoC router, interconnected via a network bridge. The PEs are connected to the DNoC, while configuration registers are linked to the CNoC. This interconnection allows full access to all network components across both NoCs.

\subsubsection{Event Router} \label{subsubsec:event_router}

\begin{figure}[h!tbp]
    \centering
    \includegraphics[width=1\linewidth]{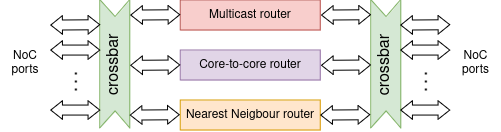}
    \caption{SpiNNaker2 event router}
    \label{fig:s2erouter}
\end{figure}

The event router is responsible for routing multicast event packets, as well as core-to-core, nearest neighbor packets and global read/write packets, shown in Fig.~\ref{fig:s2erouter}. Packets arrive from other chips via the link interfaces and NoC packets from PEs and are presented to the router through its 6 NoC ports. Packets are decoded and arbitrated by a front-end crossbar to different internal engines in parallel.

The SpiNNaker2 event router follows the SpiNNaker router concept \cite{furber2013spinnarch}, extending it via more routing entries and more flexible packet types.
It operates at a clock speed of 400~MHz.
Several packet types with different functionality have been implemented:
\begin{itemize}
    \item {\bf Multicast Packet}: This packet is provided for communicating event messages to multiple targets in a multi-chip system \cite{furber2013spinnarch}. It contains a 32~bit source ID that the router uses to forward the packet to one or multiple targets, dependent on the routing entries. Each of the 16K routing entries contains a 32~bit source filter, indicating for each bit either a match or a wildcard, and a set of output links and PEs to forward the packet to. The packet only has 40~bits and can be extended by up to 128~bits of payload, supporting transfer of scalars or short vectors together with the event.
    \item {\bf Core-to-Core Packet}: This packet is meant for sending a message from one core (PE) in a multi-chip system to one other core in the system, identified by a 16~bit chip address and an 8~bit core address. Up to 128~bit payload can be added to the packet.
    \item {\bf Nearest-Neighbour Packet}: This packet is provided for sending data to a neighbouring chip or retrieving data from it. It contains a 32~bit address and an optional payload of at most 128~bit.
    \item {\bf Global Read/Write Packet}: This packet allows direct access to any address on any chip in a multi-chip system by means of a target chip, target core and target address. Access can either be a write or a read request. Read data is sent back to the source.
\end{itemize}
The multicast packet is especially fitted for event-based communication, supporting a tree-like distribution to multiple targets and low number of bits to be transported, minimizing communication load. 
The multicast router supports a theoretical peak throughput of \SI{1843.2}{\giga\bit/\second} to PEs and \SI{460.8}{\giga\bit/\second} to external links.

The SpiNNaker2 event router provides 16 configurable hardware diagnostic counters, conceptually similar to those available in SpiNNaker. These counters enable fine-grained monitoring and classification of specific packet types based on programmable filtering criteria. Details of the internal router architecture and the counters are provided in the Supplementary Material.

\subsubsection{Interfaces}
The SpiNNaker2 chip provides interfaces to connect external memory and peripherals,
an external controller, a host computer
and to communicate with other SpiNNaker2 chips.
\begin{figure}
    \centering
    \includegraphics[width=1.0\linewidth]{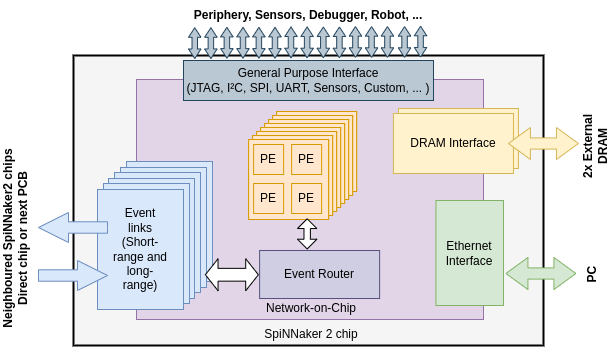}
    \caption{SpiNNaker2 communication interfaces}
    \label{fig:s2links}
\end{figure}

\paragraph{LPDDR4}
Two LPDDR4 memory interfaces
can connect external memory of up to 4\,GByte each.
The interface contains a \emph{Uniquify} LPDDR4 PHY
and controller operated at
800\,MHz clock frequency providing a raw bitrate
of \SI{25.6}{\giga\bit/\second}.
A DMA controller for each interface
can offload larger data transfers
from applications while preventing NoC congestion.

\paragraph{Event links}
Fast and reliable communication is required to transport events between chips. We separate two scenarios: on-board chip to chip connections with a distance of a few centimeters and inter-board communication with distances up to 1.5\,m.
Both types are available for each link and can be enabled mutually exclusively.
Following the different physical requirements, the links make use of different IO standards for most power efficient communication.
Each chip offers six event links to use in the hexagonal routing grid, see Fig.~\ref{fig:s2hexgrid}.

\underline{Short-range chip-to-chip links:}
Short-range communication is implemented with a Chiplet-like interface.
The eight input-output pad cells use a voltage of only \SI{0.5}{\volt} for power-efficient low-swing transmission.
Six data lanes with double-data-rate (DDR) signaling use a \SI{500}{\mega\hertz} differential clock signal (generated by a local ADPLL) with 0.25~V common mode, resulting in a transmission speed of \SI{6}{\giga\bit/\second}.
Source synchronous transmission offers advantages for power and robustness here.
Each event link has a separate physical transmitter and receiver interface that work independently and can be kept in power saving mode individually.
Data transfer makes use of an error detection with 12~bit cyclic-redundancy-check (CRC-12) \cite{7773}.
In case of data loss, an ID field in the packet is used to resend data and to ensure proper data order.
Each packet has a size of 96 bit.
It is transmitted within two system clock cycles at 125 MHz as there is a serialization factor of 8.
For power saving during link idle, clock and data signals are driven to ground, which allows immediate reactivation at next packet request.

\underline{Long-range board-to-board links:} 
Long range event communication uses a different communication standard, employing two LVDS IO pads per link, one pad per direction.
A 1 GHz transmission submits data with DDR at \SI{2}{\giga\bit/\second}, with the TX clock being generated by a local ADPLL.
The clock is recovered at the receiver with clock-data-recovery (CDR).
For ensuring the required signal toggles for CDR, the link uses 8b10b coding \cite{amd_aurora_8b10b_2014}.
As a consequence, the link needs to stay active to keep clock synchronization on the target chip, which allows power saving only with longer lead time.
Each short-range and long-range link has several packet counters to monitor event transmission. Details are described in the Supplementary Material.

\paragraph{Ethernet}

The SpiNNaker2 chip provides a \SI{1}{\giga\bit/\second} UDP Ethernet interface to connect to an external source.
Connection to an external PHY is done via SGMII, using its standard frequency of \SI{625}{\mega\hertz} with DDR.
Furthermore, to prevent data loss, we extend the UDP protocol by providing UDT support \cite{gu2007udt}. 
As a reliable UDP-based data transfer protocol, UDT adds control packets to counteract packet loss, and encompasses congestion control to avoid a transmission jam.
As a simpler alternative, we also provide an optional packet counter field after the UDP/UDT header to catch potential packet loss.
The Ethernet-to-NoC interface can interpret incoming UDP payloads with specific magic headers as single or multiple NoC packets with dynamic destinations on- or off-chip. It also provides programmable routing LUTs for specific outgoing packet types, outgoing source coordinates, incoming source ports or general raw data.    

\paragraph{GPIO Interfaces}

The SpiNNaker2 chip features 27 GPIOs operating at 1.8V, offering a wide range of connectivity options for external sensors, microcontrollers or other peripherals. 
Specifically, the chip provides:

\begin{itemize}
    \item 4 UART interfaces
    \item QSPI masters with speeds up to \SI{200}{\mega\bit/\second}
    \item QSPI slaves with speeds up to \SI{200}{\mega\bit/\second}
    \item I\textsuperscript{2}C masters with speeds up to \SI{1}{\mega\bit/\second}
    \item I\textsuperscript{2}C slave with speeds up to \SI{1}{\mega\bit/\second}
    \item 13 PWM channels
\end{itemize}
A flexible GPIO multiplexer allows customization of the GPIO functionality based on the application.

\subsubsection{Boot and Management}

A boot controller has been integrated into the periphery part of the chip, supporting a two-stage boot process: In the first stage, the chip is responsive to one of the management slave interfaces (JTAG, I\textsuperscript{2}C, SPI) for loading configuration information, selected by two bootstrap pins. Alternatively, it can boot autonomously via reading configuration data from an external SPI flash memory. All chip resources can be accessed and initialized in this stage. Once a high-speed interface has been brought up, the second stage can use it to load the application code. Boot times depend on which chip components need to be initialized and on the size of the application code, being essentially limited by the speed of the external interface used for loading. 

For helping with boot, control and background tasks, the periphery contains an ARM M4 management processor with 128 kByte memory running at 100 MHz. It is used during boot for speeding up the bring up of external interfaces that require a complex initialization sequence. It can also be used to broadcast application code to all PEs, saving loading time via the external interface.

\subsection{Power Management Concept} \label{subsubsec:power_management}

GlobalFoundries 22FDX is an FDSOI technology which allows for adaptive body biasing (ABB) to compensate for process, voltage and temperature variations for improved speed and reduced power consumption.

The SpiNNaker2 chip utilizes Racyics ABX adaptive body biasing technology with the ABB aware implementation approach from \cite{8932444} that ensures guaranteed speed for the standard cells over all characterized process-voltage-temperature (PVT) corners and in parallel reduces leakage for most efficient operation at different voltages.
There are three different power scenarios in the chip.
Components necessary for communication and initial configuration are run in a zero-body-biased (ZBB) power domain.
This \SI{0.8}{\volt} power domain is available directly after power-up and is always-on during chip operation.
Power saving mechanisms for this domain are implemented on logic level, like clock gating and multi-bit flops.

Each PE runs in a separate switchable power domain. Depending on computational requirements, one can choose between a performance mode at \SI{0.8}{\volt} and a low-power mode at \SI{0.5}{\volt}.
The PE's clock frequency is switched in parallel to utilize the higher speed of the standard cells at higher supply voltage.
As default, PEs are optimized to run with \SI{150}{\mega\hertz} clock frequency at \SI{0.5}{\volt} or with \SI{300}{\mega\hertz} at \SI{0.8}{\volt}.
In addition, the ABX engine applies a forward-body-biasing (FBB) scheme for optimized leakage at guaranteed speed targets under these ultra-low-voltage conditions \cite{hoeppner2019abb}.
Each PE can be configured individually for a specific power / speed mode, either by a global controller or by the PE itself.
This allows very fine-grained and quick adaptation for most power efficient job execution.
We have confirmed with measurements that transient IR drop due to simultaneous power mode changes of PEs does not affect operation, given sufficient supporting capacitors on the PCB. As a fallback, the hardware semaphores in each Quad unit can be used to limit the number of PEs switching at the same time.
As most of the chip area is filled with PEs, also most parts of the chip run in ABX controlled power domains.
To reduce the number of externally provided different power supplies, all digital circuits run at either \SI{0.5}{\volt} or \SI{0.8}{\volt}.

\subsection{Physical Design Implementation}
The SpiNNaker2 chip is realized with 10 independent implementation macros, assembled in a tile based design.
Our goal was to prepare a layout that can be flexibly adjusted with macros that have a common size and interface, taking the size of a Quad as reference.
All core components are implemented as multiples of this reference size.
Only boundary components have an adjusted size related to IO connectivity.
Static timing analysis ensures proper functionality in all implementation corners within each macro.
All macro interfaces are asynchronous to remove inter-macro timing relations, thus only interface components need to be timing analyzed.
This approach allows for scaling of chip size and compute power at reduced tool runtimes and complexity.
Synchronous connections are only used for scan connectivity, having relaxed speed requirements and can make use of clock inversion for relaxed hold timing requirements.
Macros of identical type are tested in parallel with parallel pattern shifting and results are evaluated on-chip automatically, reducing test time of all 266M gate equivalents significantly.
The layout size of SpiNNaker2 chip is \SI{102}{\square\mm} and includes \SI{158.6}{\mega\bit} of memory in total.
Table \ref{tab:impl_macros} shows distribution of on-chip SRAM inside the different components.

\begin{table}
    \caption{Implementation details for subcomponents}
    \centering
    \addtolength{\tabcolsep}{-0.4em}
    \begin{tabular}{cccc}
        \hline
         Component & Count & SRAM & Power Domain \\ [0.5ex] 
         \hline
         Processing element          & 152 & \SI{1}{\mega\bit}       & FBB \SI{0.5}{\volt}/\SI{0.8}{\volt} \\
         Quad unit                   & 38  &                         & ZBB \SI{0.8}{\volt}                 \\
         DRAM interface              & 2   & \SI{896}{\kilo\bit}     & ZBB \SI{0.8}{\volt}                 \\
         Event Links                 & 6   &                         & ZBB \SI{0.8}{\volt}                 \\
         Ethernet interface          & 1   & \SI{1.875}{\mega\bit}   & ZBB \SI{0.8}{\volt}                 \\
         GPIO interface / management & 1   & \SI{1}{\mega\bit}       & ZBB \SI{0.8}{\volt}                 \\
         Event router                & 1   & \SI{2.875}{\mega\bit}   & ZBB \SI{0.8}{\volt}                 \\
         Total                       &     & \SI{158.625}{\mega\bit} &                                     \\
         \hline
    \end{tabular}
    \label{tab:impl_macros}
\end{table}

\subsection{Test Board and Software Stack}

Test boards like the one used in this article (see Fig.~\ref{fig:roverdemo}) contain a STM32H743 microcontroller (STM32) in addition to the SpiNNaker2 chip and DRAM. After turning on the board, the STM32 automatically boots the SpiNNaker2 chip via SPI. Afterwards, the SpiNNaker2 chip can be controlled via the UDP Ethernet interface and is ready for application.

A software stack for SpiNNaker2 was developed consisting of three main components:
1) \emph{Chip software}: Bare metal ARM programs to run on the SpiNNaker2 PEs are written in C and compiled using a GCC toolchain with custom linker scripts. C-library functions are available for on-chip communication and control of hardware units such as accelerators. 
2) A C\textsuperscript{++}-based \emph{low-level host software} provides memory-mapped access via UDP to the chip's register files, SRAM, and DRAM. Further, there are high-level functions to load ARM programs to PEs and for memory block transfers between host and chip.
3) \emph{Python-based high-level software} for SNN and DNN applications:
\texttt{py-spinnaker2}\cite{vogginger2024pyspinnaker2} allows to define and simulate SNNs consisting of neuron populations and projections with static synapses similar to PyNN\cite{davison2009pynn}.
Inference of PyTorch or ONNX DNN models can be run using the \texttt{OctopuScheduler}\cite{langer_octopuscheduler_2025,jobst2025multilayer} framework. Only a limited number of layer types and operators is currently supported.

The \emph{SpiNNaker2 Developer Portal} at \url{https://spinnaker2.gitlab.io/} provides an overview of the latest software as well as hardware documentation.

\section{Results}

We have benchmarked the SpiNNaker2 chip in a range of different scenarios, to present a detailed energy and performance profile of the architecture.
The chip has been manufactured in GlobalFoundries 22FDX technology and put on a test board (see bottom right of Fig. \ref{fig:roverdemo}) that allows power measurement on the different voltage domains of the chip separately.
All numbers in this section are chip measurements.
In case workloads were too short for power measurement, they were repeated periodically and average power extracted.
We start with characterizing baseline power and individual accelerators and continue with spiking and deep neural network workloads, as well as more general event-based algorithms.
The code used for the measurements is available at \url{https://gitlab.com/tud-hpsn/public/s2-chip-paper}.

\subsection{Baseline Power Characterization}

We first present the results of profiling the baseline power in 2 different power modes for 3 different scenarios (see Tab. \ref{tab:chip_leakage}). 
PEs operate at their nominal clock frequencies: \SI{150}{\mega\hertz} at \SI{0.5}{\volt} supply voltage or \SI{300}{\mega\hertz} at \SI{0.8}{\volt} (see Sec. \ref{subsubsec:power_management}). If the PEs are off, power and clock are gated. The PEs are in sleep mode if the ARM M4F cores are in their low-power standby state through the WFI (Wait For Interrupt) or the WFE (Wait For Event) instructions. Multiple blocks, like the machine learning accelerator, are clock gated during sleep mode. 
To measure PE core power without active accelerators, we run the general-purpose coremark algorithm \cite{coremark} on all 152 PEs in parallel. Midst coremark execution, all ARM M4F cores and SRAM are fully active.

\begin{table}
    \caption{SpiNNaker2 chip power for idle operation and coremark processing. The first column describes the execution mode and which power lane is supplied to the PEs. The total includes power of periphery, IO and PLL.}
    \centering
    \begin{tabular}{lcccc}
        \hline
         Corner              & PE & 152 PEs & total & throughput \\
                             & [mW] & [mW] & [mW] & [CoreMark/s] \\ [0.5ex] \hline
         PEs off             & 0.4     & 59.7         & 75.2    &   \\ %
         PEs sleep@0.50V     & 1.4     & 214.7        & 235.4   &   \\ %
         PEs sleep@0.80V     & 3.6     & 544.0        & 564.7   &   \\ %
         PEs coremark@0.50V  & 2.7     & 403.9        & 424.6   &  65968 \\ %
         PEs coremark@0.80V  & 8.1     & 1227.2       & 1247.9  &  138168  \\ %
         \hline
    \end{tabular}
    \label{tab:chip_leakage}
\end{table}

\subsection{Numerical Accelerator}
\begin{figure}
    \centering
    \includegraphics[width=1.0\linewidth]{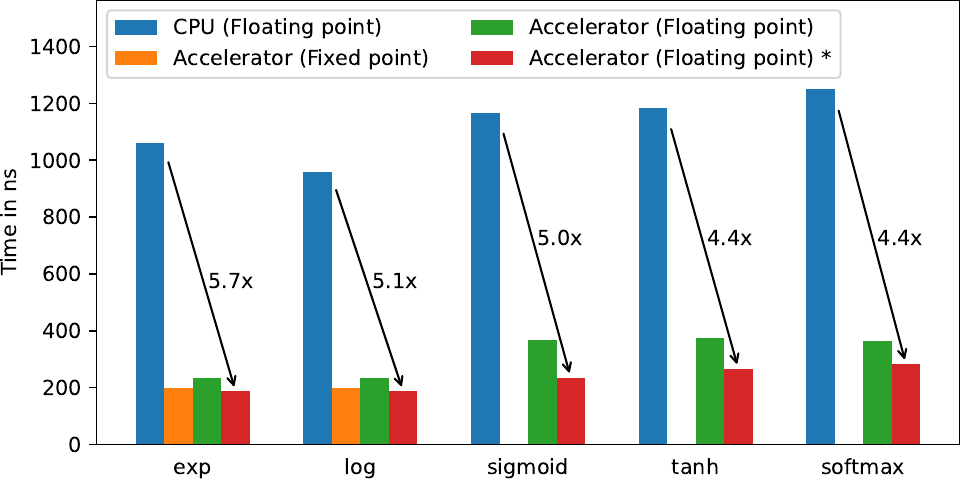}
    \caption{Time measurements for the numerical accelerator, running at \SI{150}{\mega\hertz}. ($^*$ - optimized implementation)}
    \label{fig:nmu-results}
\end{figure}

To assess the speed of the numerical accelerator with exp/log execution in software on the ARM core, we perform time measurements for the bare exponential and logarithm, as well as for the activation functions sigmoid, tanh and softmax, which heavily use the exponential function (see Fig.~\ref{fig:nmu-results}).

On the ARM core, computing one exponential using the standard C math library in single-precision floating point takes 1060\,ns (at \SI{150}{\mega\hertz} clock frequency). The accelerator uses 32\,bit fixed point formats as inputs and outputs, so we show results with 32\,bit fixed point, giving a speed-up of 5.3x and 4.8x for exponential and logarithm. For the activation functions, we convert single-precision floating-point inputs to 32\,bit fixed point before sending them to the accelerator, running at full accuracy. Afterwards, we convert the result back to floating point for the remaining computations on the ARM core. Therefore, we only observe speed-ups of 3.2x-3.5x. An implementation with optimized loop, where processing time of the accelerator is used for post-processing the previous output, improves the speed-up to between 4.4x (tanh and softmax) and 5.7x (exponential).

\subsection{Machine Learning Accelerator (MLA)}
\label{sec:mla}
We benchmark speed and energy efficiency of the machine learning accelerator (MLA) for typical DNN operations and compare it to execution on the ARM processors with the \mbox{ARM CMSIS} library \cite{lai2018cmsis}. We measure execution time and energy efficiency of 6 layers selected from ResNet-18 \cite{he_deep_2015} and an Open Pre-Trained Transformer (OPT) model \cite{zhang_opt_2022} (see supplementary material for layer details). Layers are partitioned to fit into the 128\,kB SRAM per PE and optimized for MAC cell utilization, assuming that inputs and weights are stored entirely in the on-chip SRAM.
All data was averaged over multiple measurements.
\begin{figure}
    \centering
    \includegraphics[width=1.0\linewidth]
    {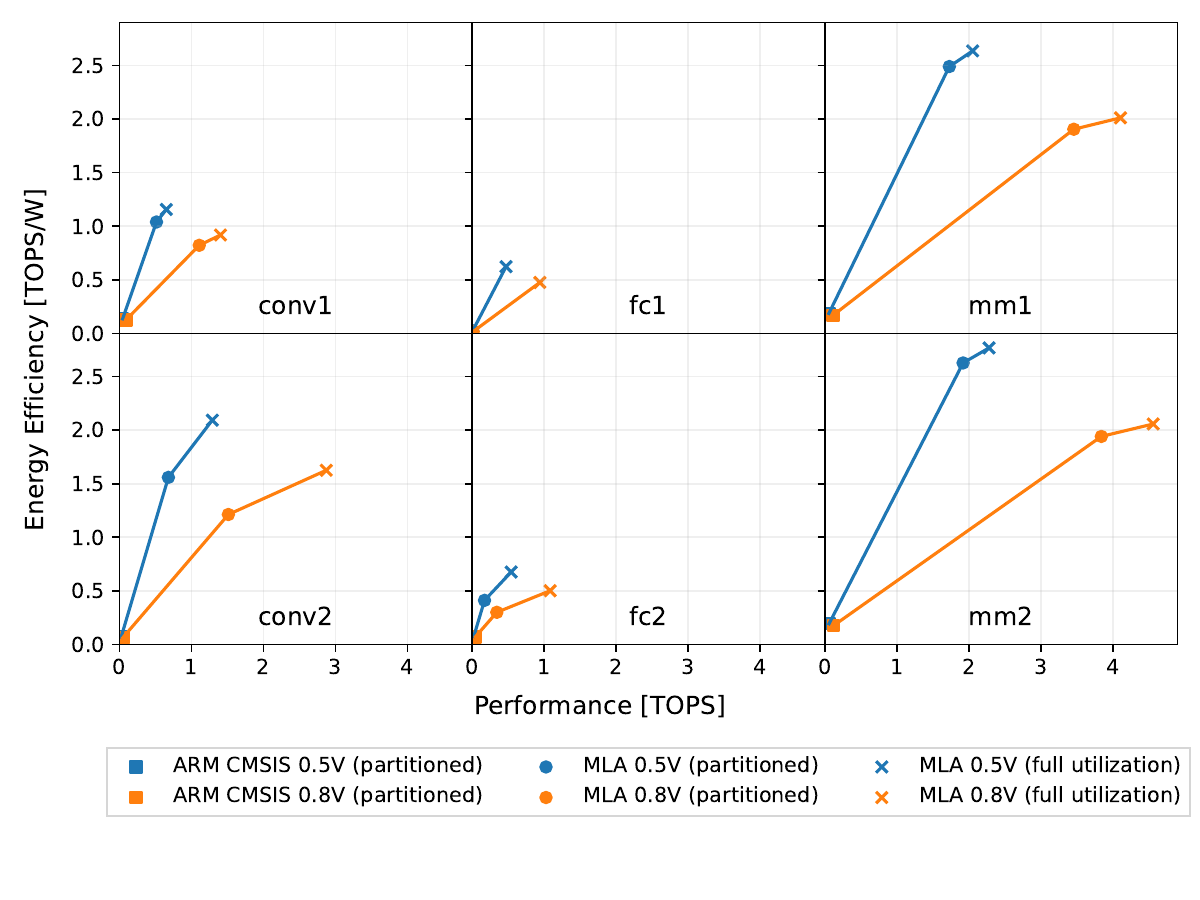}
    \caption{Performance and energy efficiency for 6 DNN layers using ARM CMSIS library versus MLA (weights stored in local PE memory). \textit{Partitioned} refers to layer execution according to a DNN tiling algorithm (see Sec. \ref{sec:dnn}), \textit{full utilization} refers to a scenario if all 152 PEs would be used with the same configuration.}
    \label{fig:armnn_vs_mla}
\end{figure}

Fig. \ref{fig:armnn_vs_mla} shows measurements of ARM core execution, MLA-accelerated execution of partitioned layer tiles and MLA-accelerated execution for full chip utilization (tiled operation on all 152 PEs) for high-throughput (\SI{300}{\mega\hertz} clock) and low-power mode (\SI{150}{\mega\hertz} clock).
Compared to computation on the ARM core, a partitioned layer execution with MLA has a significantly higher throughput (up to 34.3x / 7.6x / 32.0x for convolutional / fully-connected / matrix-matrix-multiplication layers, \SI{300}{\mega\hertz} clock) and is significantly more energy-efficient (up to 22.6x / 6.8x / 13.5x respectively, \SI{150}{\mega\hertz} clock).
For layer fc1, the MLA performance is very close to CMSIS, since only 2 out of 152 PEs are used in the partitioned version. In comparison, layer fc2 shows a more noticeable difference due to the usage of 48 PEs.
The theoretical maximum throughput of the MLA is 5.837\,TOPS, whereas practically 4.563\,TOPS (\SI{78}{\%} MAC utilization) could be achieved for matrix-matrix multiplications within a Transformer model \cite{zhang_opt_2022} for full chip utilization (\SI{300}{\mega\hertz} clock), see Fig. \ref{fig:armnn_vs_mla}.

For fully-connected layers the MAC utilization is only between \SI{16}{\%} and \SI{19}{\%} as only 1 of 4 rows of the MAC array is used.
Instead, for convolutional layers the MAC utilization is between \SI{22}{\%} and \SI{50}{\%}, showing a higher overhead for configuration and write-back of results to SRAM compared to matrix-matrix multiplication.

Although designed mainly for neuromorphic applications, SpiNNaker2's energy efficiency for deep neural network inference is comparable to the state of practice achieved by other platforms (see Tab. \ref{tab:energy_efficiency_comparison}), when operators are allocated entirely in the on-chip SRAM.

\begin{table}[]
\caption{Comparison of SpiNNaker2 against different other INT8 inference platforms for DNN}
\centering
\begin{tabular}{@{}ccccc@{}}
\toprule
Platform   & \begin{tabular}[c]{@{}c@{}}Throughput\\ {[}TOPS{]}\end{tabular} & \begin{tabular}[c]{@{}c@{}}Power\\ {[}W{]}\end{tabular} & \begin{tabular}[c]{@{}c@{}}Efficiency\\ {[}TOPS\,/\,W{]}\end{tabular} \\ \midrule
\begin{tabular}[c]{@{}c@{}}\textbf{SpiNNaker2} \textbf{(150~MHz, 0.5~V)}\end{tabular} &  \textbf{2.281\textsuperscript{*}}   & \textbf{0.825\textsuperscript{*}}   & \textbf{2.77}     \\ \begin{tabular}[c]{@{}c@{}}\textbf{SpiNNaker2} \textbf{(300~MHz, 0.8~V)}\end{tabular} &  \textbf{4.563\textsuperscript{*}}   & \textbf{2.219\textsuperscript{*}}    & \textbf{2.06}    \\ \midrule

Greenwaves GAP9\cite{reuther_lincoln_2023}          & 0.151     & 0.64          & 0.34     \\
Coral EdgeTPU\cite{reuther_lincoln_2023}            & 4         & 2             & 2.00       \\
Intel Mobileye Eye Q5\cite{reuther_lincoln_2023}    & 12        & 5             & 2.40       \\
Jetson Orin Nano\cite{nvidia_corporation_nvidia_2024}         & 67        & 25            & 2.68       \\ 
IBM NorthPole\cite{cassidy_114_2024, modha_neural_2023}           & 200       & 74            & 2.70       \\ \begin{tabular}[c]{@{}c@{}}Groq Tensor Streaming\cite{reuther_lincoln_2023}\end{tabular}  & 820       & 300           & 2.73     \\
ARM Ethos N77\cite{reuther_lincoln_2023}           & 4.1       & 0.8           & 5.13     \\ \bottomrule
\end{tabular}
\\(* denote measured values, see Fig. \ref{fig:armnn_vs_mla})
\label{tab:energy_efficiency_comparison}
\end{table}

\begin{figure}
    \centering
    \includegraphics[width=\linewidth]{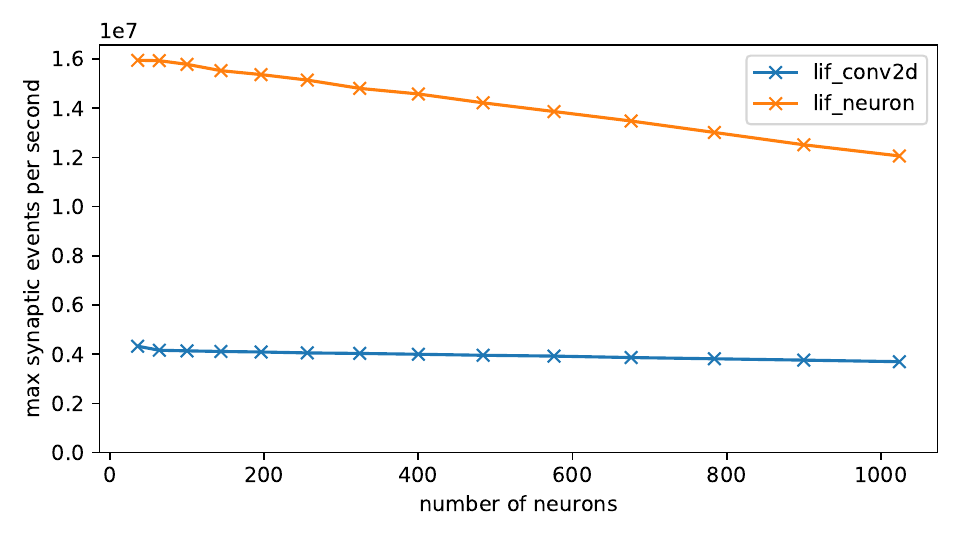}
    \caption{Maximum synaptic events per second per core (at \SI{300}{\mega\hertz} clock) at a 1 ms tick depending on number of target neurons for sparse connectivity (\texttt{lif\_neuron}) and convolutional connectivity (\texttt{lif\_conv2d}).
    }
    \label{fig:numberofneurons}
\end{figure}

\begin{figure*}
    \centering
    \includegraphics[width=0.3\linewidth]{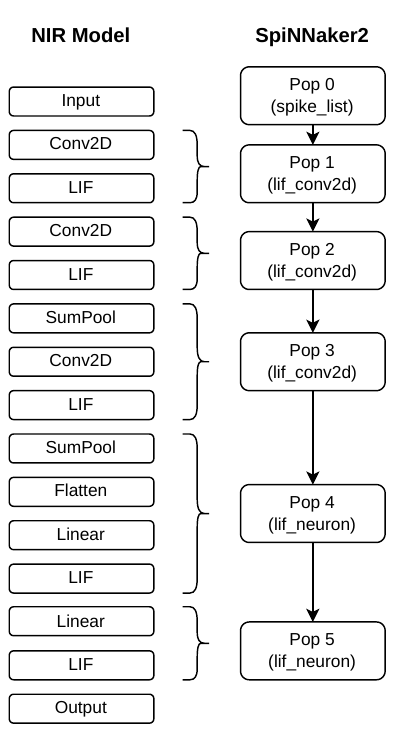}
    \includegraphics[width=0.6\linewidth]{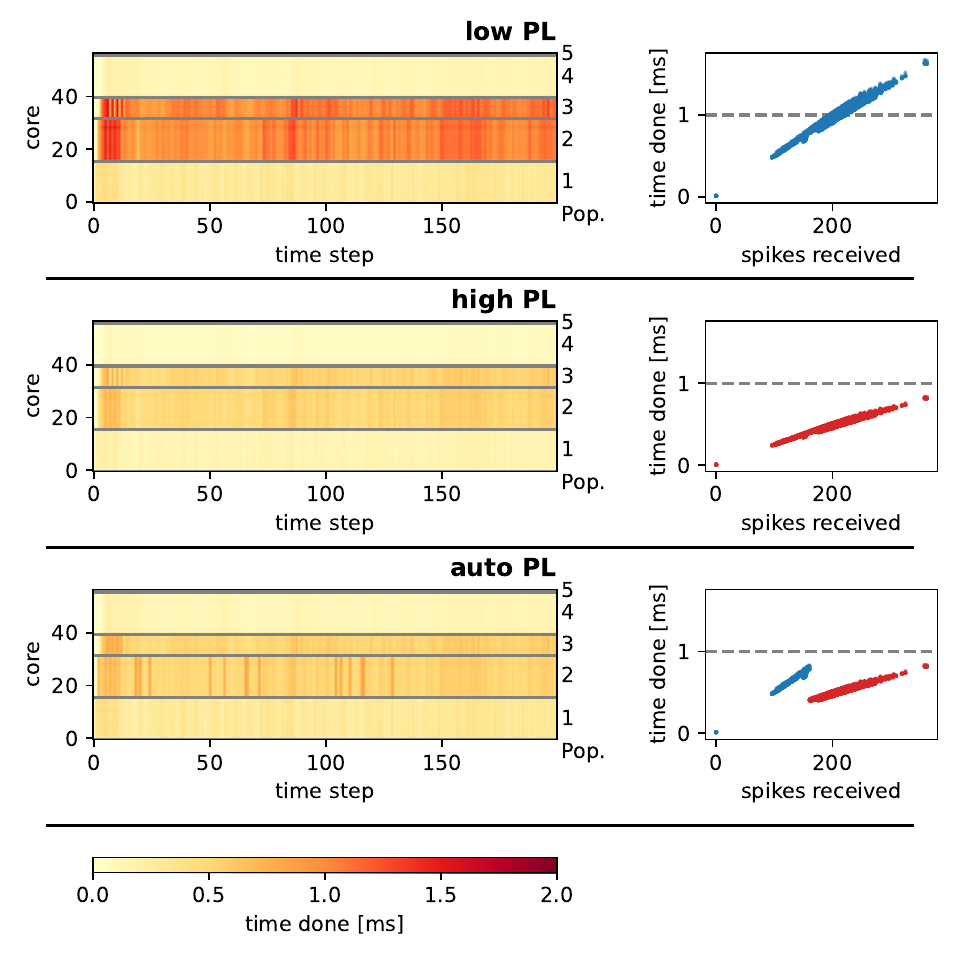}
     \caption{SNN for DVS-Gesture recognition using dynamic power management.
     Left: NIR model of deep SNN and how it is translated to populations on SpiNNaker2.
     Center: Time needed (``time done'') for processing the incoming spikes and performing neuron updates per core and simulation step.
     The rows show the results for different performance levels (PLs): low PL (\SI{0.5}{\volt}, \SI{150}{\mega\hertz}), high PL (\SI{0.8}{\volt}, \SI{300}{\mega\hertz}), auto PL (automatic selection of PL according to number of received spikes).
     Right: scatter plot of ``time done'' vs. number of received spikes per time step for population 3 (all cores, all time steps, blue dots: low PL, red dots: high PL)
     }
     \label{fig:dvfs-snn}
\end{figure*}

\subsection{Spiking Neural Networks (SNN)}
\label{sec:snn-results}
On SpiNNaker2 SNNs are executed in a similar way to SpiNNaker1 \cite{rhodes2018spynnaker}:
Each processing element simulates a number of neurons and their incoming synapses.
The neuron models, such as leaky integrate-and-fire neurons (LIF), are updated in discrete time steps using the Euler method in software running on the ARM core \cite{arfa2025efficient}.
In this work, all neuron parameters, state variables and synaptic weights are stored in the PE's SRAM, which limits the size of the networks that can be implemented.
This limitation will disappear once the DRAM is supported in the SNN software.

At each time step (e.g., every 1 ms), PEs wake up synchronously by a timer interrupt.
Each PE first processes the spikes received during the previous time step, retrieving the corresponding target neurons from the 32-bit spike ID and adding the synaptic weights to the neurons' synaptic input buffers.
Then, the neurons are updated.
When a neuron spikes, a spike packet (SpiNNaker2 multicast packet with 32-bit key) is sent via the event router to other PEs, where it is received by the event handler.
Differing from SpiNNaker1, the 32-bit keys are stored to a spike FIFO in  SRAM without interrupting the processor, see supplement for detail.
PEs process spike packets and neuron updates sequentially, hence the time needed for processing depends on the input spike rate, the connection density and the number of neurons.
Note that PEs may get out of phase if the processing demand is too high, hence exact reproducibility is not guaranteed (See supplement for details).

\paragraph{Core Capacity}
The number of neurons per core (PE) is an important metric in neuromorphic chips, as it indicates the scale of the overall system. However, this number varies significantly with parameters such as weight precision, logging, connectivity and simulation time-step. To provide a glimpse of the core capacity in the SpiNNaker2 chip, the maximum number of synaptic events that can be maintained at a 1 ms simulation tick are presented in Fig.~\ref{fig:numberofneurons} as a function of the number of neurons.
For sparse connectivity and 1024 neurons per core, up to 12 million synaptic events can be processed per second.
Conditions of the characterization are given in the Supplementary Material, which also contains a tentative comparison to SpiNNaker.
As mentioned, the trade-offs affecting the number of neurons are numerous.
Studying them is out of the scope of this paper.

\paragraph{Application Example}
As an example, we show the simulation of a deep SNN for DVS gesture recognition \cite{amir2017low} using the chip's power management feature (Sec.~\ref{subsubsec:power_management}).
In \cite{arfa2025efficient} we have demonstrated a complete end-to-end pipeline for deploying deep SNN models on the SpiNNaker2 chip.
This includes the training of the SNN in PyTorch, the conversion to the neuromorphic intermediate representation (NIR) \cite{pedersen2024neuromorphic}, and the execution on the SpiNNaker2 chip using post-training quantization.
In this work, we focus on the real-time simulation of the SNN (as if events arrive live from a DVS camera) and apply adaptive power management.
The SNN contains 5 layers with in total \num{7859} neurons which are mapped to 57 cores of the chip (additional 64 cores are used for the input spike population), see Fig.~\ref{fig:dvfs-snn} and supplement for details. %
When simulated for 1000 time steps (corresponding to 1 second of the input data) on SpiNNaker2, we achieve a test accuracy of $92.04\%$ which is less than the $94.0\%$ achieved in PyTorch for the full gestures with on average 6 seconds duration.
Longer simulations were not possible with the current software stack due limited spike storage in SRAM.

We apply and compare 3 different power management strategies for the SNN execution:
\begin{itemize}
    \item \textbf{Low PL}: In the low performance level, all cores are operated at \SI{0.5}{\volt} and \SI{150}{\mega\hertz}. 
    \item \textbf{High PL}: In the high performance level, all cores are operated at \SI{0.8}{\volt} and \SI{300}{\mega\hertz}.
    \item \textbf{Auto PL}: Here we apply dynamic voltage and frequency scaling (DVFS) as introduced in \cite{hoeppner2019dynamic} to automatically switch between the high and low performance level in each core and time step. The high PL is chosen if the number of received spikes exceeds a user defined threshold.
\end{itemize}
The results are presented in Table~\ref{tab:dvs_snn}, alongside energy measurements across the different performance levels as shown in Fig.~\ref{fig:dvfs-snn}.
When operated at low PL, we used a simulation tick of 3 ms to make sure that the neural processing finishes for all cores. In Fig.~\ref{fig:dvfs-snn} (heat map for low PL) one can see that many cores need close to 2~ms to finish the processing (``time done'').
Instead, at high PL, a simulation tick of 1~ms is sufficient to finish the processing on time due to the higher core clock frequency.
Finally, we apply the automatic PL selection: For this, we extracted from simulations the dependency of the ``time done'' and the number of received spikes per core (scatter plots in the right column of the figure) and determined a threshold spike count above which the high PL is used.
When applying auto PL, the processing always finishes within 1 ms for real-time processing but only operates at the high PL when needed.
By this, $28\%$ less energy is consumed compared to the high PL (Tab.~\ref{tab:dvs_snn}).

\begin{table}[htbp]
\caption{SNN DVS gesture prediction on SpiNNaker2.}
\label{tab:dvs_snn}
\centering
\begin{tabular}{@{}llll@{}}
\toprule
Mode                    & Low PL    & High PL & Auto PL \\ \midrule
Simulation tick  [s]    & 0.003     & 0.001  & 0.001 \\
Energy [J]              & 2.010     & 1.023  & 0.741 \\ 
Accuracy (\%)           & 92.04     & 92.04  &  92.04 \\ \bottomrule
\end{tabular}
\end{table}

Further details about the SNN model, the deployment pipeline and comparison to other works can be found in \cite{arfa2025efficient}.

\subsection{Deep Neural Networks (DNN)}
\label{sec:dnn}
To support the inference of state-of-the-art DNN architectures like convolutional neural networks (CNNs), transformer-based large language models (LLMs) or vision transformers, the scheduling framework \textit{OctopuScheduler} has been developed (see \cite{langer_octopuscheduler_2025}, \cite{jobst2025multilayer},
\cite{gonzalez2023spinnaker2}). It interprets SpiNNaker2 as an MPSoC platform, coordinating the execution of standard DNN layer types such as convolutional layers, fully connected layers, matrix-matrix-multiplications, and others. The scheduling within a layer and for a complete DNN is performed completely on-chip to minimize the communication latency within and between layers. A single \textit{scheduler PE} controls the start and termination of up to 151 \textit{worker PEs} within a DNN layer. Following an automatically derived and optimized tiling scheme, the \textit{workers} perform required DNN layer computations using the PE-local machine learning accelerator (MLA) on the partitioned input and weight tensors, each producing an output tile. The \textit{workers} generally operate independently and asynchronously within a layer on equally sized tensor tiles, avoiding intra-layer synchronization and thus minimizing the communication overhead between \textit{scheduler} and \textit{worker}. For a complete multi-layer DNN model, each layer execution is triggered synchronously by the \textit{scheduler}
(see \cite{jobst2025multilayer}).

We have benchmarked the same 6 DNN layers from Section~\ref{sec:dnn} at full size using the OctopuScheduler, which includes loading and storing of data to/from DRAM, data preprocessing like data alignment, and the actual execution of the MLA.
Execution time, power and energy for each layer are compared to server and edge GPUs (A100 and Jetson Orin Nano) in Tab~\ref{tab:dnn_benchmark}.

\begin{figure}
\centering
\begin{subfigure}{.318\linewidth}
  \centering
  \vspace{-0.5\baselineskip}
  \includegraphics[width=.95\linewidth,interpolate=false]{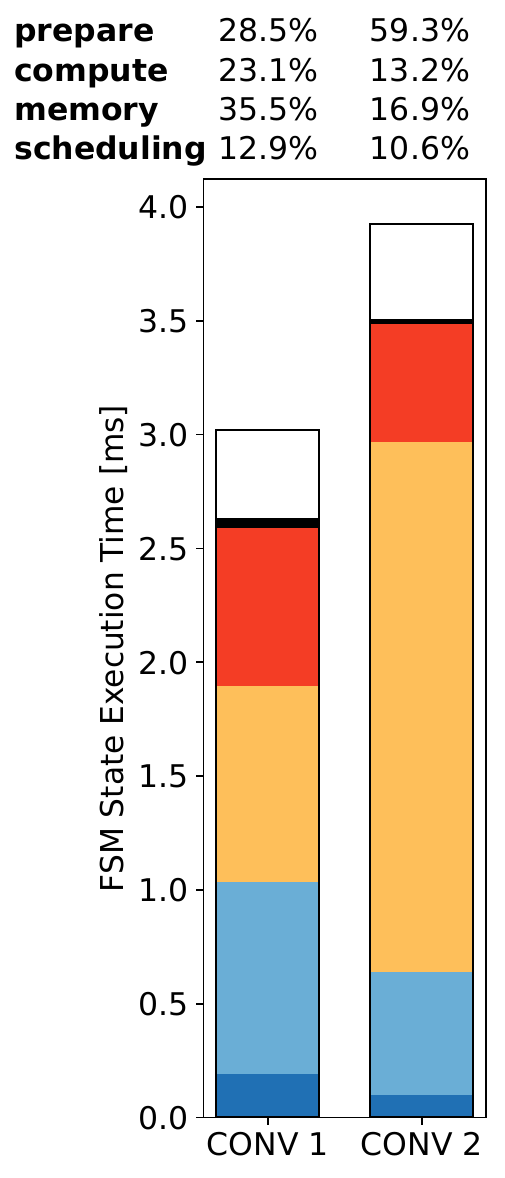}
  \vspace{-0.4\baselineskip}
  \caption{conv2d}
  \label{fig:dnn_exec_times_conv2d}
\end{subfigure}%
\begin{subfigure}{.283\linewidth}
  \centering
  \includegraphics[width=.95\linewidth,interpolate=false]{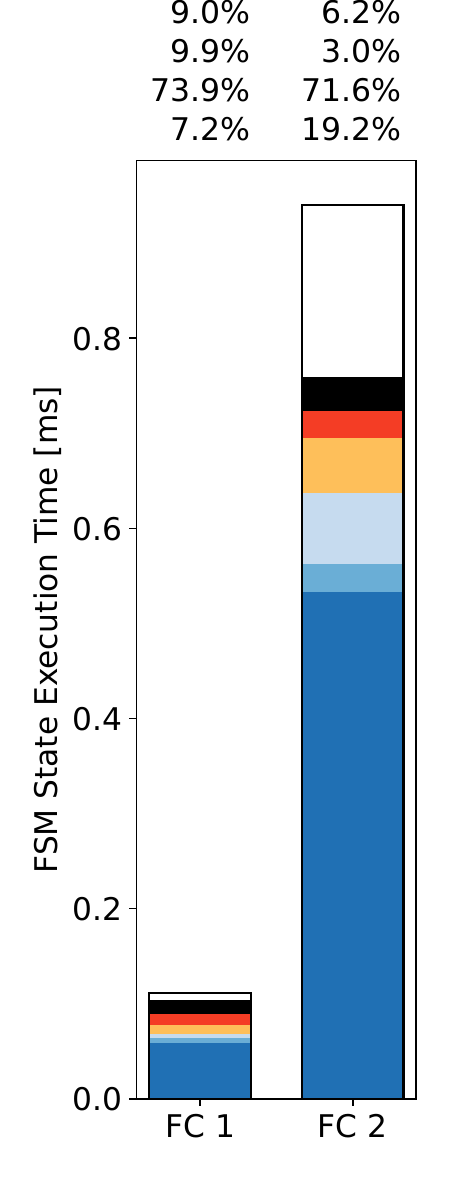}
  \vspace{-0.6\baselineskip}
  \caption{vec-mat-mul}
  \label{fig:dnn_exec_times_vec_mat_mul}
\end{subfigure}%
\begin{subfigure}{.283\linewidth}
  \centering
  \includegraphics[width=.95\linewidth,interpolate=false]{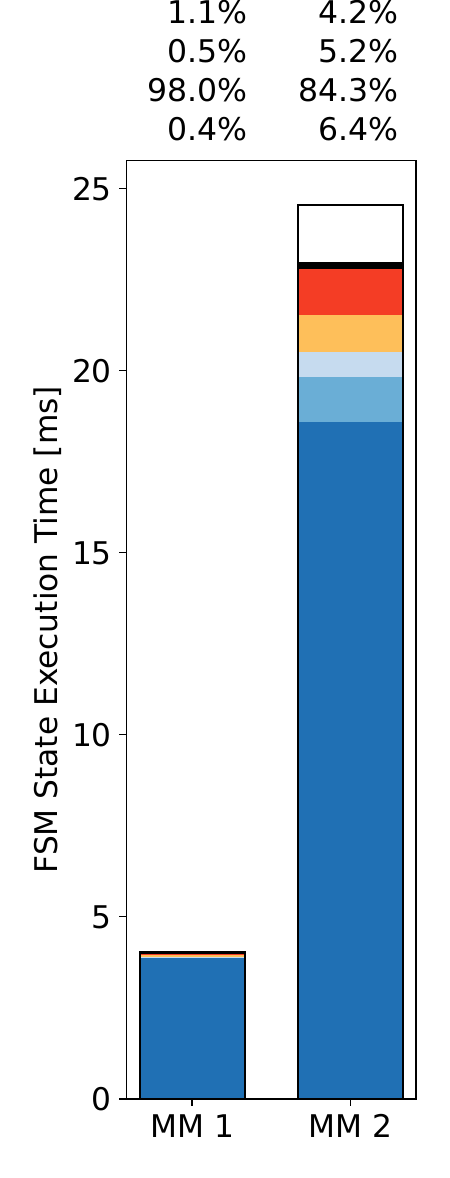}
  \vspace{-0.6\baselineskip}
  \caption{mat-mat-mul}
  \label{fig:dnn_exec_times_mat_mat_mul}
\end{subfigure}%

\begin{subfigure}{\linewidth}
  \includegraphics[width=.95\linewidth]{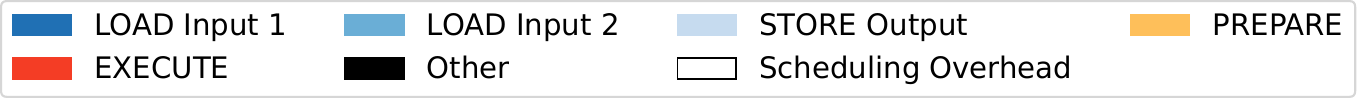}
\end{subfigure}

\caption{Execution times of layers based on (\protect\subref{fig:dnn_exec_times_conv2d}) 2D convolutions, (\protect\subref{fig:dnn_exec_times_vec_mat_mul}) vector-matrix-multiplications, and (\protect\subref{fig:dnn_exec_times_mat_mat_mul}) matrix-matrix-multiplications at \SI{150}{\mega\hertz} clock. \textit{prepare} refers to on-chip input reordering between successive layers for (a) and to MLA input alignment for (b, c), \textit{compute} to execution of the operation kernel using MLA, \textit{memory} to storage transfers of inputs and outputs, \textit{scheduling} to overhead of \textit{scheduler} execution time and synchronization compared to average \textit{worker} execution time (see proportional shares per layer at top).}
\label{fig:dnn_exec_times}
\end{figure}

As shown in Fig.~\ref{fig:dnn_exec_times}, the largest part of the matrix-multiplication layer execution time results from the DRAM transfers of input, weight or output tensors. For convolutional layers, costly off-chip DRAM transfers of intermediate activations can already be avoided by reordering intermediate activations in the on-chip SRAM between two convolutional layers, showcasing the optimization potential for matrix-multiplication layers. The heavy impact of memory transfers further suggests the use of alternative inference paradigms like pipelined execution or depth-first scheduling, which minimize off-chip communication.
The \textit{worker}-\textit{scheduler} utilization (average \textit{worker} execution time compared to complete \textit{scheduler} execution time) ranges from 80.9\% to 99.6\% for all layers, showing a small scheduling overhead.
A detailed description of the DNN scheduling framework \textit{OctopuScheduler} can be found in \cite{langer_octopuscheduler_2025}. %

\begin{table}
\caption{Measurement results of DNN layer benchmark on SpiNNaker2 (\SI{150}{\mega\hertz} clock, including DRAM transfers and scheduling overhead) compared with Nvidia A100 GPU using \texttt{torchao} and Jetson Orin Nano using \texttt{torch2trt} (all INT8).%
}
\label{tab:dnn_benchmark}
\centering
\begin{tabular}{@{}lllll@{}}
\toprule
Layer   & Platform      & Power (W)     & Time (ms)      & Energy (mJ) \\ \midrule
FC 1    & SpiNNaker2    & 0.577         & 0.111          & 0.064       \\ %
FC 1    & GPU        & 107.1         & 0.012          & 1.332       \\  %
FC 1    & Jetson        & 3.116         & 0.327          & 1.019       \\ \midrule
FC 2    & SpiNNaker2    & 0.800         & 0.944          & 0.755       \\ %
FC 2    & GPU           & 108.7         & 0.024          & 2.648        \\  %
FC 2    & Jetson        & 3.707         & 0.403          & 1.494       \\  \midrule
MM 1    & SpiNNaker2    & 0.612          & 3.971        & 2.429      \\ %
MM 1    & GPU           & 108.0         & 0.015          & 1.590        \\  %
MM 1    & Jetson        & 3.358         & 0.347          & 1.165       \\ \midrule
MM 2    & SpiNNaker2    & 0.697         & 24.501        & 17.073      \\ %
MM 2    & GPU           & 123.2         & 0.034          & 4.176       \\  %
MM 2    & Jetson        & 8.211         & 0.974          & 7.998       \\  \midrule
CONV 1  & SpiNNaker2    & 0.723         & 3.035          & 2.195       \\
CONV 1  & GPU           & 133.1         & 0.053          & 7.041        \\  %
CONV 1  & Jetson        & 5.834         & 0.657          & 3.833       \\ \midrule
CONV 2  & SpiNNaker2    & 0.597         & 3.918          & 2.339       \\
CONV 2  & GPU           & 116.7         & 0.039          & 4.599        \\  %
CONV 2  & Jetson        & 4.269         & 0.602          & 2.570       \\ \bottomrule
\end{tabular}
\end{table}

\subsection{Event-Based Algorithms}
\label{sec:event_algo}

In the following, we present recently developed event-based algorithms that go beyond classical spiking or deep neural networks, combining the best of both worlds.
\cite{subramoney2023} introduced an event-based version of the gated recurrent unit (GRU), applying a biologically inspired thresholding mechanism to reduce the communication between neurons. \cite{khan2024} demonstrates an implementation of the EGRU network on SpiNNaker2 for language modeling and DVS gesture recognition.
This implementation demonstrates that significant gains in energy efficiency can be reached versus conventional hardware for single-batch inference:
Energy per inference for a language model reduced by a factor of 18 compared to a GPU implementation (65~mJ on SpiNNaker2 vs. 1.19~J on Nvidia A100), with the drawback of an 8x longer execution time.

Due to its flexibility, SpiNNaker2 is ideally suited to explore novel learning algorithms:
\cite{bena2025event} used SpiNNaker2 to train SNNs on-chip with event-based backpropagation (EventProp), which is hard or even impossible on other neuromorphic systems with dedicated digital cores, but easy to realize with SpiNNaker2 as it can send events with flexible payload, e.g., for sending error signals.
For the EventProp example in \cite{bena2025event}, SpiNNaker2 required only 31\% of the energy per training step compared to a RTX 4070 GPU.
In a similar fashion, \cite{rostamiEpropSpiNNakerExploring2022} applied the biological plausible E-prop learning rule \cite{bellecSolutionLearningDilemma2020} to train recurrent SNN for keyword spotting on a SpiNNaker2 FPGA prototype.
We ported the E-prop code to the SpiNNaker2 chip and measured runtime and power: while SpiNNaker2 is significantly slower than a NVIDIA V100 GPU, it requires $8\times$ less energy when considering device utilization, see supplementary section~\ref{sec:eprop-power} for details.
Recent neuromorphic algorithms for combinatorial optimization also show promising results on SpiNNaker2 \cite{chen2025ising}.
These examples demonstrate how the flexible compute substrate of SpiNNaker2 allows for implementation of new models and for assessing their gains in energy efficiency.

\subsection{Integration with Sensors}

\begin{figure}
    \centering
    \includegraphics[width=1.0\linewidth]{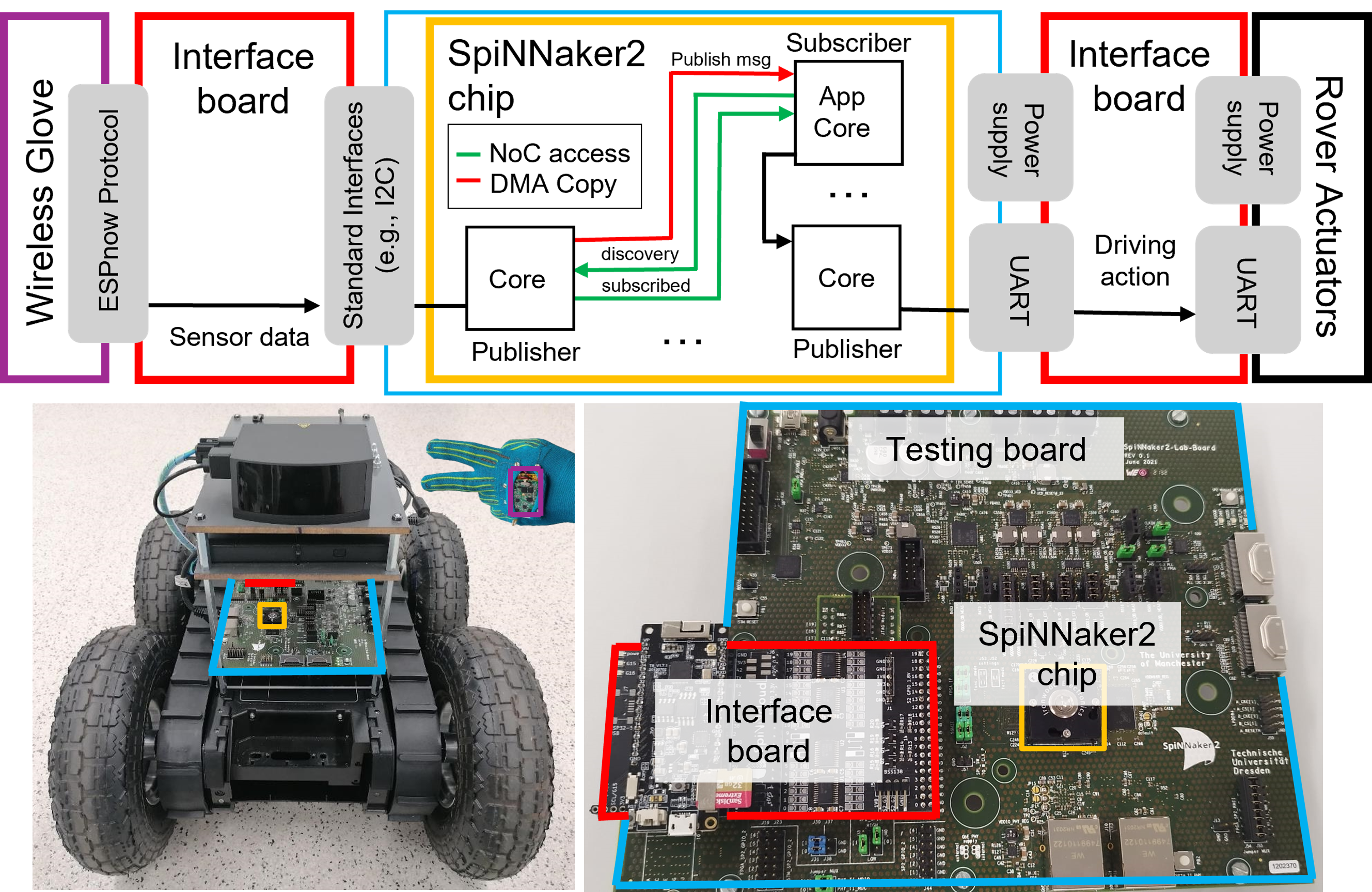}\vspace{0.3cm}\\
    \includegraphics[width=0.4\linewidth, angle=90]{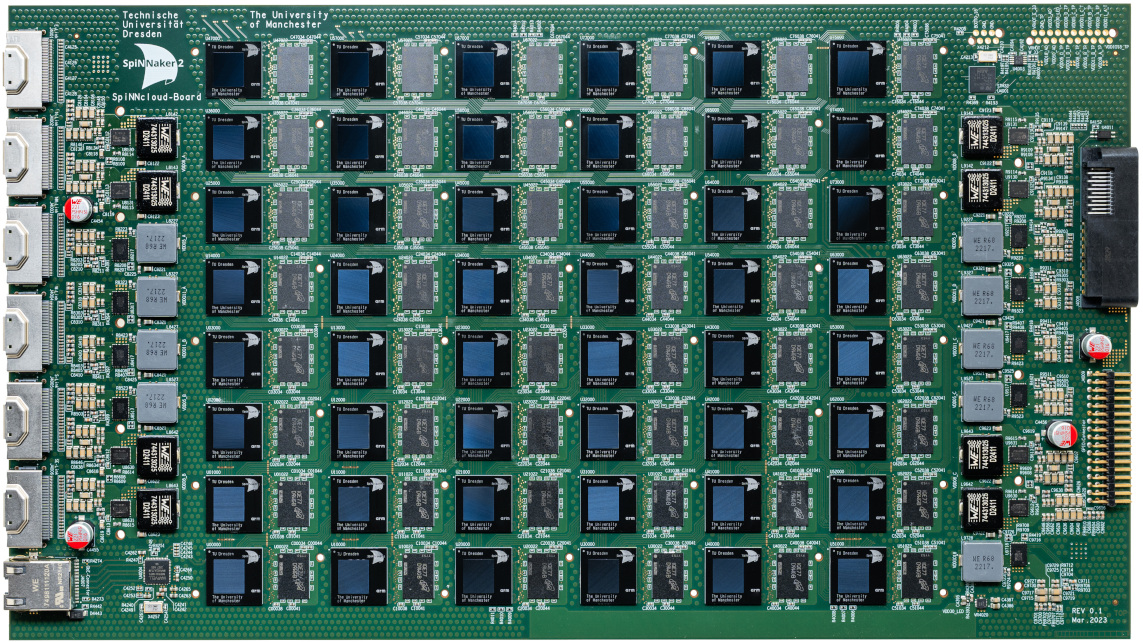}
    \includegraphics[width=0.68\linewidth]{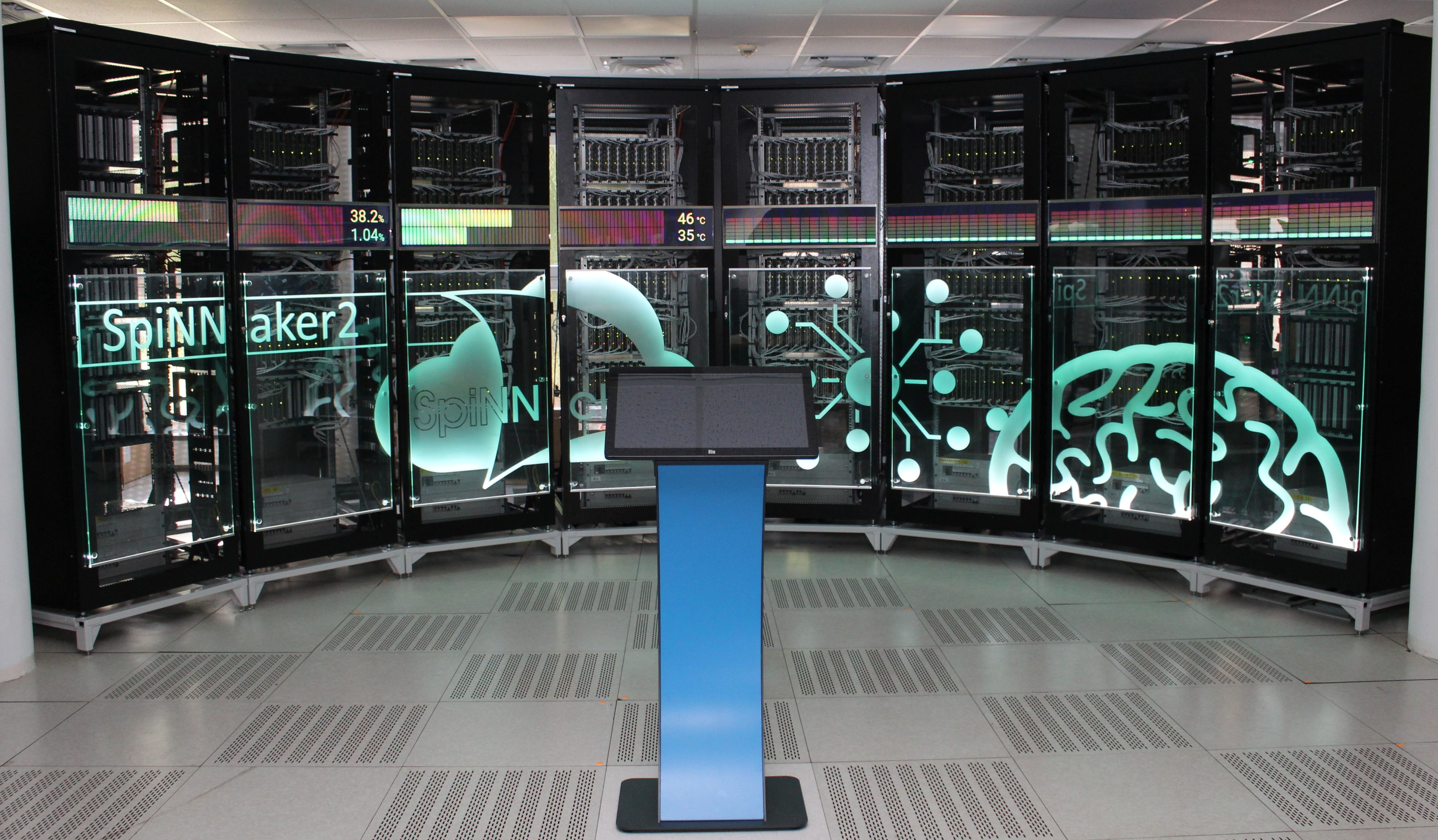}
    \caption{Top: Robotic demonstrator with SpiNNaker2, bottom: mainframe system ``SpiNNcloud'' with 48-chip board (left) and full installation (right).}
    \label{fig:roverdemo}
\end{figure}

A single SpiNNaker2 chip can also operate as a standalone system, where its internal PEs interface with external sensors and actuators. The top of Fig.~\ref{fig:roverdemo} illustrates a single-chip demonstrator, where the testing board serves as the computing platform for a four-wheel-drive rover, which also features a MEMS-based Robosense RS-LiDAR-M1 mounted on top. The example uses one core for parsing sensor data and sending driving commands via UART to the rover, another core for buffering and preprocessing, and a third core that maps driving commands from gestures acquired through a wireless glove with a GRU-based classifier, deployed with a MicroTVM-based flow \cite{liu2024gestureTVM}. The wireless glove interfaces with a daughter board using the ``ESPnow'' protocol, which forwards data to SpiNNaker2 via I\textsuperscript{2}C.

\section{Conclusion}

\begin{table*}
\caption{Comparison of SpiNNaker2 with SpiNNaker and Loihi/Loihi2
}
\label{tab:neuro_hw_comparison}
\centering
\begin{tabular}{@{}lp{0.18\textwidth}p{0.18\textwidth}p{0.18\textwidth}p{0.18\textwidth}@{}}
\toprule
                 & Loihi      & Loihi2     & SpiNNaker      & {\bf SpiNNaker2} \\ \midrule
Technology node  & 14~nm   & 7~nm         & 130~nm          & 22~nm   \\
Clock frequency  & variable     & 1000~Mhz         & 180~MHz    & 150/300~MHz    \\
Area  &60~mm$^2$ & 60~mm$^2$ & 102~mm$^2$ & 102~mm$^2$ \\
On-chip SRAM     & 33~MByte & 31~MByte & 1.8~MByte & 19.8~MByte \\
Power            & $<1.5$~W & 1.5-2.5~W & 0.36-1~W & 0.24-2.2~W \\
Number of cores  & 128 & 128 & 18 & 152\\
Core features    & Neuro-synaptic core with programmable data pipeline & Neuro-synaptic core with programmable data pipeline, graded spikes, convolution support & ARM M0 processor & ARM M4F processor, FP32 support, exp/log, RNG and GMM/convolution accelerators \\
Learning support & STDP, three-factor learning & STDP, three-factor learning & Software-based & Software-based \\
Routing          & mesh core-to-core spike routing & mesh core-to-core spike routing & Core-to-core and multicast & Core-to-core and multicast, NoC for on-chip data transfer\\
Spiking neuron capacity & \num{131072} & \num{1048576} & \num{16000} & \num{150000}\\
\bottomrule
\end{tabular}
\end{table*}

In this paper, we have presented the neuromorphic many-core chip SpiNNaker2.
We introduced its architecture and its main components and features.
Finally, we demonstrated its performance for a variety of computational workloads, including spiking neural networks, deep neural networks, as well as event-based computing approaches.

While the focus in this paper was on single-chip results, the SpiNNaker2 system architecture has been designed for a wide range of system sizes.
As shown in Fig. \ref{fig:roverdemo}, it is used both in small-scale demonstrators, as well as cloud-scale systems, with the current installation at TU Dresden reaching a size of 5 million cores \cite{kudithipudi2025neuromorphic}, with 35k chips in eight racks.
The software stack for large-scale deployment on this system is currently under development.

As shown in Tab. \ref{tab:neuro_hw_comparison}, SpiNNaker2 increases capacity by almost an order of magnitude compared to its predecessor SpiNNaker, both in terms of on-chip memory and core count, as well as spiking neuron capacity.
Moreover, the hardware accelerators available in the SpiNNaker2 PEs can further increase neuron capacity significantly \cite{huang2023efficient}, which has not been incorporated in the comparison.
Moreover, the numerical accelerators (exp, log, RNG) and the native floating point support in the processors facilitate implementation of learning algorithms and improve throughput, e.g. speeding up event-based weight updates for STDP learning rules via fast exponential computation. Additionally, more flexible packet types simplify communication of learning signals.
We are currently working on NESTML \cite{linssen2025nestml} support, which will facilitate studies on a great variety of learning models.

The hardware accelerators in each PE also enable DNN and event-based algorithms to run efficiently on the chip and thus widen the flexibility and applicability of the system.
Still, SpiNNaker2 is not a pure DNN inference accelerator.
While efficiency for local DNN processing is comparable to more DNN-centered hardware platforms (see Tab. \ref{tab:energy_efficiency_comparison}), limited DRAM bandwidth often reduces utilization of the DNN accelerators if DNN layers are too big to be stored in on-chip memory (see Fig. \ref{fig:dnn_exec_times}).
Thus, scalability of conventional DNN workloads on a single chip is limited, except for applications where DNN layers stay small and massive parallelization over the input data is required.
Bigger DNN models could be spread onto multiple chips via mapping different layers onto different chips and pipelining their execution. By this, DNN size per chip could be reduced to allow more on-chip data re-use and limit DRAM data transfer requirements. Moreover, depth-first execution could be utilized to avoid buffering of activations in DRAM between layers.
We have started exploring these options.
Still, we expect that more sparse and event-based deep neural network algorithms will result in higher efficiency improvements on SpiNNaker2 (see Sec.~\ref{sec:event_algo}).
The Loihi/Loihi2 family of systems targets a similar application range, while differing in its architecture from SpiNNaker/SpiNNaker2 by dedicated hardware blocks for synapses and neurons. Together with implementation in a smaller technology node, this allows a higher number of neurons per area. On the other hand, the range of supported algorithms is restricted on Loihi/Loihi2 to SNN and extensions towards event-based algorithms using graded spikes. The software-based approach of the SpiNNaker/SpiNNaker2 systems offers greater flexibility. Moreover, SpiNNaker2 also offers more routing possibilities, with high-speed data transfer on-chip and transmission of data alongside event packets between chips.

The variety of workloads demonstrates the increased flexibility of the SpiNNaker2 chip for distributed computing with time-varying computing demands.
This opens up possibilities for efficiently realizing new types of machine learning algorithms that reduce overall computing and memory demands by allowing more irregular compute and communication patterns.
That approach is orthogonal to GPU and classical deep learning accelerators, which achieve their high efficiency from large regular and predictable memory, compute and communication workloads.
Saving computations on such hardware platforms on a fine-grained level often does not pay off in terms of energy or performance.
In contrast, the SpiNNaker2 chip better allows capitalization on these kind of savings via its low baseline power and hardware support for workload adaptation.
At the same time, the SpiNNaker2 chip is designed for real-time operation and integration with sensors and actuators, making it highly suitable for robotics applications.

While there are specific hardware blocks for e.g. deep networks or event handling, the chip is suited for any application that can be distributed onto a large number of tiny compute elements that exchange small chunks of data.
We are looking forward to explorations of such novel, more efficient algorithms with the SpiNNaker2 chip.

\bibliographystyle{ieeetr}
\bibliography{spinnaker2}

\clearpage

\section{Supplementary Material for the paper "The SpiNNaker2 chip: a many-core platform for flexible and scalable brain-inspired computing"}

\subsection{Internal Structure of Event Router}

\paragraph{Router Architecture}

\begin{figure}
    \centering
    \includegraphics[width=1.0\linewidth]
    {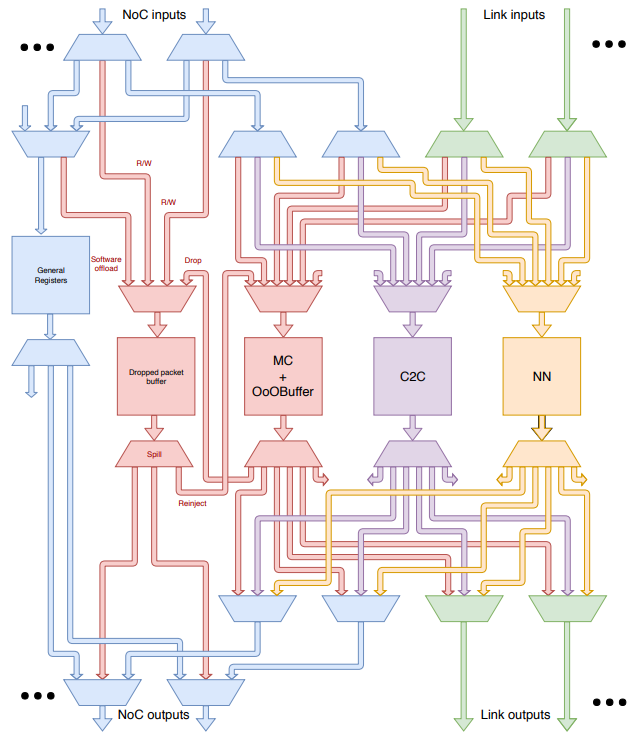}
    \caption{Internal structure of the SpiNNaker2 event router}
    \label{fig:router_details}
\end{figure}

The architecture of the SpiNNaker2 event router is shown in Fig.~\ref{fig:router_details}.
The router occupies the silicon area of two Quad Processing Elements (QPEs) and is connected to six bidirectional Network-on-Chip (NoC) ports. This configuration enables simultaneous packet reception and transmission across all six ports, providing high aggregate bandwidth.

Incoming packets from the six input ports are first processed by an input crossbar interconnect, which performs arbitration and distributes packets to three routing engines operating in parallel. This design enables concurrent processing of heterogeneous packet types, increasing overall throughput and reducing input blocking. Configuration packets are routed directly to the register file module.

Each routing engine can independently forward packets to different output ports. Consequently, multiple packets can be routed and issued in parallel within the same cycle, provided that output port conflicts do not occur. This architecture minimizes structural hazards and sustains high throughput under balanced traffic.

\paragraph{Multicast Routing Engine}

Multicast packets are handled by a dedicated multicast routing engine implementing a four-stage pipeline:

\begin{enumerate}
    \item TCAM Lookup Stage: Performs associative matching of packet address field against routing entries.
    \item Priority Encoding Stage: Resolves multiple matches and selects the highest-priority routing entry.
    \item Link-Destination Resolution Stage: Determines the set of 7 off-chip links.
    \item Core-Destination Lookup Stage: Determines internal 152 core destinations.
\end{enumerate}

To improve energy efficiency, the fourth stage can be dynamically disabled when no internal core destinations are present, reducing unnecessary lookup activity and lowering dynamic power consumption.

\paragraph{Output Scheduling and Contention Handling}

To mitigate output contention and avoid pipeline stalls, an out-of-order issue buffer is implemented at the MC routing engine output.

If a packet cannot be transmitted due to a blocked output port, it is temporarily stored in the issue buffer rather than stalling the entire pipeline. The scheduler continuously monitors output availability and selects the earliest packet that becomes eligible for transmission.

This mechanism provides reduced output blocking, improved output port utilization, higher sustained throughput under contention, resulting in lower average packet latency compared to strictly in-order issue

\subsection{Diagnostic Counters for Event Routing}

As mentioned in the main text, the SpiNNaker2 event router provides 16 configurable hardware diagnostic counters.  Each 32-bit diagnostic counter is associated with a configurable filter control register, allowing selective packet counting according to the following header attributes:
\begin{enumerate}
    \item Destination Field (8 bits), distinguishing between the seven off-chip links, the monitor core, other cores and dropped packets.
    \item Source Field (2 bits), differentiating between off-chip packets and on-chip packets originating from processing elements (PEs).
    \item Payload Size Field (4 bits), encoding different packet lengths: packets without payload, packets with 32-bit payload, 64-bit payload or 128-bit payload.
    \item Routing Algorithm Field (2 bits), indicating whether default routing was applied or a non-default routing scheme was used.
    \item Packet Type Field (3 bits), classifying packets as Nearest Neighbour (NN), Core-to-Core (C2C) or Multicast (MC) packets.
\end{enumerate}

In addition to packet classification counters, the SpiNNaker2 router integrates several dedicated hardware profiling counters to support performance analysis and debugging:
\begin{enumerate}
    \item Router Cycle Counter: Measures the number of clock cycles consumed during a specified router operation.
    \item Busy Cycle Counter: Counts router wait cycles, i.e., cycles during which backpressure occurs at any of the six router input ports.
    \item Zero-Wait Packet Counter: Records the number of packets that traverse one of the MC, C2C, or NN routing engine outputs without incurring any delay.
    \item Iterative Drop Counter: Counts packets that are dropped more than once during routing.
    \item Error Packet Counter: Records packets associated with routing errors, including time-phase violations and unroutable packet errors.
    \item Total Reinserted Packet Counter: Counts the number of packets that are reinserted into the routing fabric by hardware mechanisms.
    \item Dropped packet Water-Level Counter: Tracks the number of SpiNNaker packets currently stored in the dropped-packet buffer. The associated water-level register records the maximum occupancy reached in this buffer.
\end{enumerate}

\subsection{Diagnostic Counters for Event Links}

Each short-range and long-range link has four diagnostic counters for monitoring transmission.
The four counters are incremented at the following conditions:
\begin{enumerate}
    \item Error in CRC of received packet
    \item Correct sequence ID in received packet
    \item Link partner requested re-send
    \item Wrong header in received packet
\end{enumerate}
Each counter is 8~bit wide.

\subsection{Event Handler and Spike FIFO}
As mentioned in Section~III.B.1 and Fig.~\ref{fig:s2pe}, each PE has an event handler to process incoming SpiNNaker packets without interrupting the processor.
The event handler has two filters and one default handler.
Each filter, also called ``spDMA'' allows to automatically transfer the content of received SpiNNaker2 packets to an SRAM-mapped FIFO.
Each filter is configured with 1 bit per packet type (Multicast, core-to-core or nearest-neighbour) and 1 bit per payload size. There is a match if the type and payload bit corresponding to the received packet are enabled. In addition, one can configure what to store: the control field, the key-field, and payload.
The selected data is stored into a hardware FIFO mapped to SRAM. The FIFO size can be set by specifying the start and end SRAM addresses.
User code can access and modify the current read and write pointers, read the current fill-level and bits indicating whether the FIFO is full or empty.
If the FIFO is full, new packets are either dropped and counted, overwrite existing data, or stall the NoC interface (depending on configuration).
While the FIFO is not full, there is no back-pressure on the NoC. 
Interrupts to the ARM processor can be generated for various cases such as the FIFO being non-empty, having reached a certain fill-level or being full.
The default handler works in the same way but processes all packets not previously matched by filter 1 or 2.

\subsection{Spiking Neural Networks}
\subsubsection{Synchronization}
The synchronization of PEs for SNNs works as follows:
\begin{enumerate}
    \item For each PE, the ARM program is loaded and started to call setup functions according to configuration data stored in SRAM. Then, the PE goes into WFI (wait for interrupt) state.
    \item A chip-global interrupt wakes up all PEs simultaneously. Then, the PEs start the timer to trigger a timer interrupt every $N$ reference clock cycles, for example $N=1000$ for a \SI{1}{\ms} time step considering the \SI{1}{\mega\hertz} reference clock, cf.~Section~III.B.1.  Again, the PEs go into WFI state.
    \item The timer interrupt wakes up all PEs: The neuron cores first read the current spike FIFO's read and write pointer position to know which spikes were received in the previous time step. Then, the program starts to process these spikes, performs synapse and neuron updates, sends out spikes, and records spikes and voltages (if enabled). Afterwards, the PEs go into WFI state.
\end{enumerate}
If in 3) the next timer interrupt is issued before the processing of the previous time step finished, first, the processing of the previous step will run until completion. The processing of the next time step will start delayed so that the spike FIFO may contain spikes received in the new time step, which however will be treated as being from the previous step. Hence, the simulation may get out of phase. Yet, PEs can catch up in subsequent time steps to get in synch again.
Such incidents are logged during simulation and reported to the user afterwards.

Multi-chip SNN simulation will use the so-called SCAMP program running on each chip to trigger the chip-global interrupt. The SCAMP programs on different chips and boards use a software mechanism to measure and compensate for their relative drift \cite{rhodes2020real}.

\subsubsection{SNN characterization}
\label{appendix:snn-characterization}
Details for the study in Fig.~\ref{fig:numberofneurons}:
We used the high-level library for SNN  \texttt{py-spinnaker2} \cite{vogginger2024pyspinnaker2} for the characterization.
Two different neuron models are used:
\texttt{lif\_neuron} -- leaky integrate and fire (LIF) neuron with sparse incoming connections (individual synapses stored as list-of-lists similar to sPyNNaker \cite{rhodes2018spynnaker}), and \texttt{lif\_conv2d} -- LIF neuron with 2D convolutional connectivity using shared weights.
A \texttt{spike\_list} population is used to provide input spikes. 
The experiment is set up such that the number of synaptic events equal for both neuron models.
The 2D convolution has 1 input layer, 4 output channels and a 3x3 kernel.
The size of each output channel is varied between 3x3 (36 neurons) up to 16x16 (1024 neurons).
Each spike arriving in one core leads to 4x3x3=36 synaptic events.
Then, the number of input spikes per time steps is sweeped and the "time done" is measured. The maximum number of timesteps is then interpolated.
The sweep for the sparse connectivity uses the same neuron numbers as for the 2D convolution, as well as the same number of synaptic events per incoming spikes (36).

\subsubsection{SNN application}
For the Deep SNN in section~IV.D we used the following partitioning and placement strategy of populations to SpiNNaker2 PEs:
In \texttt{py-spinnaker2}, we manually defined the maximum number of neurons per core (PE) for each population as given in Tab.~\ref{tab:snn-partitioning}. The software automatically splits larger populations accordingly into multiple population slices and maps them to separate cores on the SpiNNaker2 chip.
A simple, linear placement is used: Populations and their slices are placed in provided order to the PEs sorted by Quad coordinate and PE id inside the quad.
Here, the number of neurons per core per population was manually adjusted considering memory constraints and to balance computational load.

\begin{table}
    \caption{Partitioning of the Deep SNN for DVS-Gesture recognition to SpiNNaker2 PEs (cores)}
    \label{tab:snn-partitioning}
    \centering
    \begin{tabular}{|c|c|c|c|}
        \hline
        \textbf{Population} & \textbf{Neurons} & \textbf{Neurons per Core} & \textbf{Nr. of Cores}\\
        \hline
        0 (\texttt{spike\_list})& 4096 & 32 & 64\\
        1  (\texttt{lif\_conv2d})& 3600 & 225 & 16\\
        2  (\texttt{lif\_conv2d})& 3600 & 225 & 16\\
        3  (\texttt{lif\_conv2d})& 392 & 49 & 8 \\
        4  (\texttt{lif\_neuron})& 256 & 16 & 16 \\
        5  (\texttt{lif\_neuron})& 11 & 11 & 1 \\
        \hline
    \end{tabular}
\end{table}

\subsubsection{Comparison to SpiNNaker}
While it would be desirable to provide quantitative numbers on the advancements of SpiNNaker2 over SpiNNaker, a fair comparison of the neuron and synapse processing capacity is currently not possible due to different software stacks.

A similar capacity study like the one shown in Fig.~9 has been performed in \cite{knight2016synapse} for SpiNNaker.
They report maximum \num{6.144} million synaptic events per second and core for 256 leaky integrate-and-fire neurons using \SI{100}{\%} connection density.
Instead, on SpiNNaker2 we achieve \num{15.14} million synaptic events at \SI{14}{\%} connection density with 256 neurons per core. Here, SpiNNaker2 is roughly $2.5\times$ better than SpiNNaker.

While both approaches use the same format (list of lists) for representing sparse connection matrices, there are the following major differences between implementations:
\begin{itemize}
    \item \textbf{Clock frequency}: \SI{300}{\mega\hertz} on SpiNNaker2 vs.\ \SI{200}{\mega\hertz} for SpiNNaker.
    \item \textbf{Neuron model}: A simplified leaky integrate-and-fire neuron without synaptic time constants (SpiNNaker2) vs.\ slightly more complex leaky integrate-and-fire neuron with exponential synapses (SpiNNaker)
    \item \textbf{Synapses}: 16-bit synapses stored in SRAM (SpiNNaker2) vs.\ 32-bit synapses stored in SDRAM (SpiNNaker)
    \item \textbf{Neuron updates}: float32 (SpiNNaker2) vs.\ 32-bit fixed-point (SpiNNaker)
\end{itemize}
This makes a fair comparison impossible. Instead, we look forward to a future direct comparison between the two systems once the \texttt{sPyNNaker} \cite{rhodes2018spynnaker} has been migrated to SpiNNaker2.

We also highlight additional work for SpiNNaker splitting synapse and neuron processing to different cores, thereby achieving a higher throughput of synaptic events, see \cite{knight2016synapse}, \cite{rhodes2020real}, and \cite{peres2022parallelization}.

\subsection{DNN Layer Parameters}
\label{sec:dnn_layer_params}
Tables \ref{tab:dnn_layer_params_conv} and \ref{tab:dnn_layer_params_fc+mm} contain the DNN layer parameters taken from prototype layers of ResNet-18 \cite{he_deep_2015} and an Open Pre-Trained Transformer (OPT) model \cite{zhang_opt_2022}. The prototype layers were used for the benchmark of MLA against ARM CMSIS execution (see \mbox{section \ref{sec:mla}} and \mbox{Fig. \ref{fig:armnn_vs_mla}}), focusing only on the layer computation under the assumption that all layer weights and inputs are present locally in SRAM. Additionally, they were used for the full DNN layer benchmark (incl. storage transfers) within the \textit{OctopuScheduler} framework (see \mbox{section \ref{sec:dnn}}, \mbox{Fig. \ref{fig:dnn_exec_times}}, and Tab. \ref{tab:dnn_benchmark}). The \textit{partitioned layer parameters} refer to a distributed layer execution according to the tiling algorithm of \textit{OctopuScheduler} \cite{langer_octopuscheduler_2025}.

\begin{table}[h]
\caption{Parameters for complete and partitioned 2D-convolutional (CONV) layers implemented with \textit{OctopuScheduler}}
\label{tab:dnn_layer_params_conv}

\begin{subtable}{1\linewidth}
\subcaption{complete CONV layer parameters}
\centering
\addtolength{\tabcolsep}{-0.2em}
\begin{tabular}{|c|ccccccccc|}
\hline
\multirow{2}{*}{\textbf{\begin{tabular}[c]{@{}c@{}}Layer\\ Name\end{tabular}}} & \multicolumn{3}{c|}{\textbf{Input}}    & \multicolumn{3}{c|}{\textbf{Weights}}  & \multicolumn{3}{c|}{\textbf{Output}} \\ 
      & C  & H   & \multicolumn{1}{c|}{W}   & H & W & \multicolumn{1}{c|}{Strides} & C  & H   & W   \\ \hline
CONV 1 & 3  & 240 & \multicolumn{1}{c|}{320} & 7 & 7 & \multicolumn{1}{c|}{(2, 2)}  & 64 & 120 & 160 \\
CONV 2 & 64 & 60  & \multicolumn{1}{c|}{80}  & 3 & 3 & \multicolumn{1}{c|}{(1, 1)}  & 64 & 60  & 80  \\ \hline
\end{tabular}
\label{tab:dnn_layer_params_conv_full}
\end{subtable}

\vspace{\baselineskip}

\begin{subtable}{1\linewidth}
\subcaption{partitioned CONV layer parameters}
\centering
\addtolength{\tabcolsep}{-0.2em}
\begin{tabular}{|c|cccccccccc|}
\hline
\multirow{2}{*}{\textbf{\begin{tabular}[c]{@{}c@{}}Layer\\ Name\end{tabular}}} & \multicolumn{3}{c|}{\textbf{Input}}    & \multicolumn{3}{c|}{\textbf{Weights}}  & \multicolumn{3}{c|}{\textbf{Output}} & \multirow{2}{*}{\textbf{\begin{tabular}[c]{@{}c@{}}Worker\\ PEs\end{tabular}}}               \\
      & C  & H  & \multicolumn{1}{c|}{W}   & H & W & \multicolumn{1}{c|}{Strides} & C  & H  & \multicolumn{1}{c|}{W}      &    \\ \hline
CONV 1 & 3  & 13 & \multicolumn{1}{c|}{166} & 7 & 7 & \multicolumn{1}{c|}{(2, 2)}  & 32 & 4  & \multicolumn{1}{c|}{80}    & 120 \\
CONV 2 & 64 & 14 & \multicolumn{1}{c|}{82}  & 3 & 3 & \multicolumn{1}{c|}{(1, 1)}  & 4  & 12 & \multicolumn{1}{c|}{80}     & 80  \\ \hline
\end{tabular}

\label{tab:dnn_layer_params_conv_part}
\end{subtable}
\end{table}

\begin{table}[h]
\caption{Parameters for complete and partitioned fully-connected (FC) and matrix-matrix-multiplication (MM) layers implemented with \textit{OctopuScheduler}}
\label{tab:dnn_layer_params_fc+mm}

\begin{subtable}{1\linewidth}
\subcaption{complete FC and MM layer parameters}
\label{tab:dnn_layer_params_fc+mm_full}
\centering
\begin{tabular}{|c|cccccc|}
\hline
\multirow{2}{*}{\textbf{\begin{tabular}[c]{@{}c@{}}Layer\\ Name\end{tabular}}} & \multicolumn{2}{c|}{\textbf{Input}}    & \multicolumn{2}{c|}{\textbf{Weights}}    & \multicolumn{2}{c|}{\textbf{Output}}    \\
        & H   & \multicolumn{1}{c|}{W}    & H    & \multicolumn{1}{c|}{W}   & H     & W     \\ \hline
FC 1     & 1   & \multicolumn{1}{c|}{512}  & 512  & \multicolumn{1}{c|}{64}  & 1     & 64    \\
FC 2     & 1   & \multicolumn{1}{c|}{3072} & 3072 & \multicolumn{1}{c|}{768} & 1     & 768   \\ \hline
MM 1     & 512 & \multicolumn{1}{c|}{512}  & 512  & \multicolumn{1}{c|}{64}  & 512   & 64    \\
MM 2     & 512 & \multicolumn{1}{c|}{3072} & 3072 & \multicolumn{1}{c|}{768} & 512   & 768   \\ \hline
\end{tabular}
\end{subtable}

\vspace{\baselineskip}

\begin{subtable}{1\linewidth}
\subcaption{partitioned FC and MM layer parameters}
\label{tab:dnn_layer_params_fc+mm_part}
\centering
\begin{tabular}{|c|ccccccc|}
\hline
\multirow{2}{*}{\textbf{\begin{tabular}[c]{@{}c@{}}Layer\\ Name\end{tabular}}} & \multicolumn{2}{c|}{\textbf{Input}}    & \multicolumn{2}{c|}{\textbf{Weights}}    & \multicolumn{2}{c|}{\textbf{Output}}  &   \multirow{2}{*}{\textbf{\begin{tabular}[c]{@{}c@{}}Worker\\ PEs\end{tabular}}}\\
        & H  & \multicolumn{1}{c|}{W}    & H    & \multicolumn{1}{c|}{W}  & H   & \multicolumn{1}{c|}{W}     &        \\ \hline
FC 1     & 1  & \multicolumn{1}{c|}{512}  & 512  & \multicolumn{1}{c|}{32} & 1   & \multicolumn{1}{c|}{32}    & 2      \\
FC 2     & 1  & \multicolumn{1}{c|}{3072} & 3072 & \multicolumn{1}{c|}{16} & 1   & \multicolumn{1}{c|}{16}    & 48     \\ \hline
MM 1     & 16 & \multicolumn{1}{c|}{512}  & 512  & \multicolumn{1}{c|}{16} & 16  & \multicolumn{1}{c|}{16}    & 128    \\
MM 2     & 16 & \multicolumn{1}{c|}{3072} & 3072 & \multicolumn{1}{c|}{16} & 16  & \multicolumn{1}{c|}{16}    & 128    \\ \hline
\end{tabular}
\end{subtable}

\end{table}

The DNN layers were benchmarked on an Nvidia A100-SXM4-40GB GPU with CUDA 12.8 using \texttt{torchao} (version: 0.16.0). For fully connected layers, \texttt{torchao} did not support a batch size of $N\leq16$ due to internal kernel limitations (see issue \#2376), so the minimal batch size of $N=17$ was selected. Similarly, \textit{OctopuScheduler} runs fully-connected layers always for batch size $N=4\,m$ with $m \in \mathbb{N}$ due to hardware dimensions. On Jetson Orin Nano, the DNN layers were benchmarked using \texttt{torch\_tensorrt} without any restrictions.

\subsection{E-prop Characterization}
\label{sec:eprop-power}
The code from \cite{rostami2022eprop} for training a recurrent spiking neural network for keyword spotting (Google Speech Commands \cite{wardenSpeechCommandsDataset2018}) with E-prop was ported to the SpiNNaker2 chip.
Power and time measurements are shown in Table~\ref{tab:eprop} and compared to the training with an NVIDIA V100 GPU.

To cope with the low utilization of the SpiNNaker2 PEs (only 12 of 152 used), we calculate the \emph{effective power} and \emph{effective energy} which only consider the \emph{utilization fraction} of the baseline power:
\begin{align}
    P_\mathrm{effective} &= \mathrm{utilization} \times P_\mathrm{baseline} + P_\mathrm{dynamic} \\
    E_\mathrm{effective} &= P_\mathrm{effective} \times T,
\end{align}
where $T$ is the full training time.

We note that the same method as in \cite{rostami2022eprop} is used that streams the input data from a host computer to a ping-pong buffer on SpiNNaker2. We expect that the training could be significantly accelerated by storing data in DRAM and by moving to data-parallel training as done for EventProp \cite{bena2025event}. This, however, is out of scope of this paper.

\begin{table}[h!]
\caption{E-prop characterization}
\label{tab:eprop}
\centering
\begin{tabular}{|l|l|l|} 
\hline
\textbf{Measurement}                          & \textbf{NVIDIA V100} & \textbf{SpiNNaker2}  \\ 
\hline
\textbf{Batch size}                           & 100                  & 1                                 \\
\textbf{Utilization [\%]}                     & 27                   & 7.9                               \\
\textbf{Baseline Power [W]}                   & 44                   & 0.28558                           \\
\textbf{Dynamic Power [W]}                    & 47.3                 & 0.0063                           \\
\textbf{Effective Power [W]}                  & 59.18                & 0.02886                           \\
\textbf{Time [h:m]}                           & 1:58                 & 500:00                            \\
\textbf{Effective Energy [kJ]}                & 419.0                & 52.0                              \\
\hline
\end{tabular}
\end{table}

\end{document}